\documentclass[a4paper,11pt]{article}
\usepackage{jheppub} 
\usepackage{xcolor}
\usepackage[autostyle]{csquotes}
\usepackage{caption}
\usepackage{subcaption}

\newcommand{\comments}[1]{}

\title{\boldmath CMB Injection Bounds on Moduli Fields}

\preprint{LITP-26-21; {USTC-ICTS/PCFT-26-62}}

\author[a]{Naman Agarwal}
\author[a,b]{Andrew R.~Frey,}
\author[c,d]{Ratul Mahanta,}
\author[e,f,g]{Anshuman Maharana,}
\author[h,i]{Fernando Quevedo,}
\author[i]{Gonzalo Villa}
\affiliation[a]{Department of Physics \& Astronomy, \\
University of Manitoba, Winnipeg, Manitoba R3T 2N2, Canada}
\affiliation[b]{Department of Physics and Winnipeg Institute for Theoretical Physics,\\ University of Winnipeg, Winnipeg MB, R3B 2E9, Canada}

\affiliation[c]{Interdisciplinary Center for Theoretical Study, University of Science and Technology of China, Hefei, Anhui 230026, China}
\affiliation[d]{Peng Huanwu Center for Fundamental Theory, Hefei, Anhui 230026, China}

\affiliation[e]{Harish-Chandra Research Institute, \\   Chhatnag Road, Jhunsi,
Prayagraj, Uttar Pradesh 211019, India}

\affiliation[f]{ Homi Bhabha National Institute, \\ Training School Complex, Anushakti Nagar, Mumbai
400094, India}

\affiliation[g]{
 Leinweber Institute of Theoretical Physics, \\ Department of Physics, University of Michigan, Ann Arbor, MI 48104, USA
}

\affiliation[h]{New York University Abu Dhabi, Saadiyat Island, Abu Dhabi, UAE}
\affiliation[i]{DAMTP, University of Cambridge, Wilberforce Road, Cambridge, CB3 0WA, UK}

\emailAdd{agarwaln@myumanitoba.ca}\emailAdd{a.frey@uwinnipeg.ca}
\emailAdd{ratul.mahanta@ustc.edu.cn}
\emailAdd{anshumanmaharana@hri.res.in}
\emailAdd{fq2054@nyu.edu}
\emailAdd{gv297@cam.ac.uk}
\abstract{ Light scalar fields with gravitational strength interactions \textit{(moduli)} are  ubiquitous
in string, supergravity, and extra-dimensional models. 
If produced in the early universe with lifetimes exceeding the epoch of recombination, slowly decaying moduli inject energetic particles into the Standard Model plasma, leaving distinct imprints on the cosmic microwave background (CMB).
In this paper we systematically derive CMB injection bounds on light moduli ($10^3\ {\rm eV}\leq m_{\rm{mod}}\leq 10^8\ {\rm eV}$) across the primary cosmological production channels: vacuum misalignment, heavy scalar/inflaton decay, thermal emission from the primordial plasma including production from a Hagedorn phase of hot strings. We show that these injection bounds impose stringent constraints that are complementary to both the standard cosmological moduli problem and fifth force bounds.}

\begin{document}
\maketitle
\flushbottom

\section{Introduction}\label{sec: Intro}

Precise observations of the cosmic microwave background (CMB) \cite{Planck:2018vyg} have  revolutionized cosmology. The fractional contributions of radiation,
matter and dark energy to the energy budget of the universe are now
known to remarkable accuracy. Furthermore, inhomogeneities in the CMB on the superhorizon scales have provided us with a window to the very early universe. CMB data have also been important for constraining 
the basic constituents of the universe, including strong limits on the number $\Delta N_{\rm eff}$
of additional neutrino-like species.\footnote{The quantity is also independently constrained by big bang nucleosynthesis.} 

In addition to probing standard cosmological components, CMB precision measurements place stringent injection bounds  on non-standard energy release  into the  Standard Model (SM) plasma, during or after recombination, from sources such as dark matter annihilation, unstable particle decays, primordial black hole evaporation, etc.\cite{Padmanabhan:2005es, Galli:2009zc, Slatyer:2009yq, Slatyer:2012yq, Poulin:2016nat,Simon:2022ftd,  Liu:2023fgu, Xu:2024vdn}

The goal of this paper is to systematically evaluate CMB injection bounds on moduli fields: fields with gravitational-strength interactions that arise naturally in string, supergravity, and extra-dimensional models (see e.g.~\cite{Cicoli:2023opf} for a comprehensive discussion on how moduli fields can impact cosmology). The energy injected into the SM plasma during decoupling is determined by the
decay width and number density (at the time). The decay width of any modulus is set by its mass; on the other hand, the number density is determined by the production mechanism. Thus, the injection bounds depend on the production mechanism.  We derive injection bounds across the primary cosmological mechanisms for moduli production in the early universe: the vacuum  misalignment mechanism, production via the decay of the inflaton, thermal emission from the hot SM plasma, and production during an early Hagedorn phase dominated by hot strings.

In a realistic scenario, several production mechanisms are likely to take place, with one dominating over the others. In the context of moduli fields, one of the standard production mechanisms  is vacuum misalignment.  Moduli with initial displacement can lead to the cosmological moduli problem (CMP) ~\cite{Coughlan:1983ci, Banks:1993en, deCarlos:1993wie, Kawasaki:2000en}.   The initial displacement of a modulus is set by early universe dynamics and hence model dependent, but generic considerations give the initial displacement to be Planckian. For such displacements, a modulus will dominate the energy density of the universe and thereby pose an obstruction for successful nucleosynthesis (which requires radiation domination). The solution requires that such moduli decay sufficiently early and reheat the universe to at least a few MeV. This corresponds to an approximate lower bound of $30$ TeV on moduli masses.

A similar bound also applies to axions: for an axion of mass $m_a$, with axion-photon coupling $g_{a \gamma \gamma}$, and an initial misalignment\footnote{For axions, the generic expectation for the initial displacement is of the order of the  associated Peccei-Quinn scale: $f_{\rm a}$ (see e.g. \cite{Preskill:1982cy}).} so that it dominates the energy density of the universe  -- successful big bang nucleosynthesis requires $m_{a} \gtrsim  { 30 \over (g_{a \gamma \gamma} M_p)^{2/3}} \  {\rm {TeV}}$. Since the initial displacement is model dependent, it is  important to consider displacements away from generic expectations -- in regimes where (pseudo) scalars never dominate the energy density of the universe. For instance, this was done in the study of irreducible  axion backgrounds~\cite{Balazs:2022tjl,Langhoff:2022bij}; we will adopt the same logic for moduli in this article. We thus find that, even in situations where moduli are not subject to the CMP (say, due to a tuning of their initial conditions) there is a mass range that remains constrained by independent CMB considerations.

We will find that the mass range relevant for CMB injection bounds is, conservatively, $10^3\ {\rm eV}\leq m_{\rm{mod}}\leq 10^8\ {\rm eV}$, as we discuss this in detail in section
\ref{sec:review}. Thus, the analysis of this paper is for moduli that have lifetimes longer than the age of the universe and slowly inject energy into the SM plasma at the time of decoupling. Depending on their initial displacement, however, these fields may eventually dominate the energy density of the universe and are therefore subject to overclosure bounds.
We find regions of parameter space (displacement, mass) where CMB injection has more constraining power than overclosure.
Thus, the present study provides bounds in regimes complementary to that of the CMP and overclosure bounds.

An irreducible production mechanism is thermal emission, or freeze-in, which has a particularly interesting property.
Because moduli couple to the SM plasma through non-renormalisable interactions, this production mechanism is most efficient at the highest energies, where the moduli within the mass range of interest are effectively relativistic. The moduli abundance is therefore sensitive to the reheating temperature, with stronger bounds for higher reheating temperatures.\footnote{A similar phenomenon occurs for gravitational waves, where graviton freeze-in leads to the cosmic gravitational microwave background (CGMB) \cite{Ghiglieri:2015nfa, Ghiglieri:2020mhm, Ringwald:2020ist}.}
Injection bounds and other probes have been studied in the context of axion-like particles produced via freeze-in~\cite{Balazs:2022tjl,Langhoff:2022bij} (and recently in the context of the string axiverse~\cite{Cheng:2025cmb,Yin:2025amn}), which constrain a region of parameter space involving the axion coupling constant and its mass. A strength of applying this logic for moduli is that moduli interaction rates are fixed to be of Planckian strength, which allows us to bound their mass as a function of the reheating temperature only.

Lastly, we compare production from weakly coupled quantum field theories to the case of a Hagedorn phase, where the energy density of the universe is dominated by highly excited fundamental strings, following~\cite{Frey:2005jk, Frey:2023khe, Frey:2024jqy}. In this case the energy density redshifts like matter and the number density of moduli is produced throughout the whole phase, so the total number produced depends on the density of strings at the end of the Hagedorn phase. As discussed in~\cite{Frey:2005jk,Frey:2024jqy}, we assume that the open string degrees of freedom include the SM, so the continuous transition to radiation reheats the SM. Our estimate for moduli production from hot strings therefore depends on what we consider the end of the Hagedorn phase, and its comparison to production from the SM plasma also depends on the final SM reheating temperature. This situation emphasizes the importance of the continuous transition period between Hagedorn strings and radiation in determining signals of string physics.

This paper is structured as follows. Section \ref{sec:review} reviews and develops some basic material needed to apply CMB injection bounds to cosmological moduli. Section \ref{sec:productionandbounds} obtains the injection bounds, going through various production mechanisms. We conclude in \ref{sec: Conclusion}. Some technical details related to moduli production from long strings are provided in the appendix.


\section{CMB Injection bounds and Moduli}
\label{sec:review}

In this section, we provide a brief review of CMB injection bounds, collect some essential facts about moduli and develop some basic material relevant for applying CMB injection bounds to moduli.

During the epochs of recombination and decoupling, the decay (or annihilation) of metastable species into energetic photons, electrons, and other Standard Model particles injects energy into the photon-baryon fluid. The extra energy can modify the time at which decoupling occurs. This is true even for species with lifetimes much larger than the timescale of decoupling because the rare decays inject energy into the photon-baryon plasma. However from the precise measurements of the CMB, the time of last scattering is known to a high level of precision. Using this, one can constrain the energy injected into the photon-baryon plasma and impose bounds on the lifetime of metastable species as a function of mass of the species and their cosmic abundance. These are called injection bounds and have been studied in detail phenomenologically \cite{Padmanabhan:2005es, Galli:2009zc, Slatyer:2009yq, Slatyer:2012yq, Poulin:2016nat,Simon:2022ftd,  Liu:2023fgu, Xu:2024vdn}. The injection bounds depend only weakly on the mass of the decaying particle (less than an order of magnitude over a wide mass range).
Ignoring the weak dependence of the bound on the mass, it is possible to parametrize the constraint using the empirical formula 
\begin{align}\label{eq: injectionbound}
t_{\rm{decay}}>10^{25}\frac{h^2\Omega_{\mathrm{mod}}}{h^2\Omega_{\rm DM}}\,\mathrm{sec}\sim 4\times 10^{67}\frac{1}{M_{p}}\frac{h^2\Omega_{\mathrm{mod}}}{h^2\Omega_{\rm DM}} ,
\end{align}
where $h^2\Omega_{\textrm{mod}}/h^2\Omega_{\textrm{DM}}$ is the ratio of the energy densities of the metastable species and dark matter at recombination\footnote{We have used the subscript ``mod" for the decaying particle preemptively, as we will be studying
the bounds for moduli fields.} (essentially the same as the ratio today for longer-lived moduli) and $M_p$ is the reduced Planck mass. This constraint is applicable when:
\begin{itemize}
\item  The fractional density of the decaying dark matter exceeds $10^{-10}$
\begin{equation}
\label{fraccons}
  \Omega_{\mathrm{mod}}\gtrsim 10^{-10}\Omega_{\rm DM}
\end{equation}
\item  The mass of the metastable species is in the range 
\begin{equation}
\label{massrange}
10^{3} \ \textrm{eV} < m_{\rm mod} <10^{10} \ \textrm{eV} .
\end{equation}
\end{itemize}
For a smaller abundance $\Omega_{\mathrm{mod}}$, even a decay with lifetime equal
to the decoupling time injects too little energy to constrain, so this abundance places a lower limit on the injection bounds.
On the other hand, injection bounds do apply outside the mass range \eqref{massrange}, but
the right hand side of \eqref{eq: injectionbound} starts depending on the 
mass of the metastable particle.
In this article, we will conservatively confine the discussion to the parameter space described above.

The constraint on the lifetimes of these species can easily be of the order or exceed the current age of the universe ($t_0=4.35 \times 10^{17} \textrm{sec}\sim  7 \times 10^{32} \ \text{eV}^{-1}$).  
Even a species whose lifetime is of order $t_0$ has a stringent abundance bound of $h^2\Omega_{\textrm{mod}}<4\times 10^{-8}h^2\Omega_{\textrm{DM}}$. Alternatively, for a modulus to constitute the entirety of the dark matter, its lifetime must exceed the current age of the universe by at least eight orders of magnitude. 
In other words, cosmologically long-lived moduli must either have an exceptionally long lifetime or make up a very small fraction of the universal energy budget.  

Moduli fields are ubiquitous in supergravity, string and extra-dimensional models. In string compactifications (and extra-dimensional models), many of them arise from degrees of freedom associated with the shape and size of the extra dimensions (i.e., the extra dimensional metric). This origin implies that they couple to matter with couplings of gravitational strength (Planck suppressed dimension five operators).\footnote{We will refer to any species with gravitational strength interactions  as a modulus.} Their  decay width is
\begin{align}\label{eq: lifetime1}
    \Gamma_{\rm decay} \sim {m_{\rm mod}^3}\Large{/}(16\pi {M_{p}^2}) ,
\end{align}
up to order one factors.
Note that the decay width of a modulus is set by its mass. This dependence implies
that our bounds will be directly on the mass (in contrast with other particles where other
microscopic parameters would enter; for instance the decay constant, mass and axion-photon couplings for axion like particles).

Importantly, Eq.~\eqref{eq: lifetime1} implies that a modulus is stable on cosmological timescales if 
\begin{equation}
\label{eq:light}
  m_{\rm mod} < M_{p} \left({16 \pi \over M_p t_0}\right)^{1/3} \sim 3 \times 10^{-20} M_{p}
  \sim 7\times 10^{7}  \textrm{eV} .
\end{equation}
Such moduli, if produced  in the early universe,
would slowly inject energy into the Standard Model at the time of decoupling of the CMB. This makes them subject to the injection bounds.
Combining \eqref{eq:light} with \eqref{massrange} (for the modulus lifetime to be greater than roughly the time of decoupling), we arrive at the mass range for our bounds on moduli
\begin{equation}
\label{mrangefinal}
10^{3} \ \text{eV}  \lesssim  m_{\rm mod} \lesssim 10^{8} \ \text{eV}.
\end{equation}

As emphasized in the introduction, injection bounds are sensitive to the abundance
of the decaying species; this in turn depends
on the production mechanism. Next, we turn to the mechanisms by which moduli
can be produced and obtain the corresponding CMB injection bounds.

\section{Production Channels and Bounds}
\label{sec:productionandbounds}

  The two primary modes for moduli production in the early universe are
vacuum misalignment and production from decays and emissions.

\subsection{Vacuum Misalignment}

   Inflationary dynamics (more generally scalar field dynamics
in the early universe) leads to vacuum misalignment for
moduli as well as axionic fields. The field remains pinned at its
 displaced value until the Hubble constant becomes of the
 order of the mass of the modulus. Thereafter, the field oscillates about its minimum and can be described as 
 a pressure-less fluid of cold non-relativistic moduli particles.

    Consider a modulus of mass $m_{\rm mod}$  
with initial displacement $\phi_{\rm in} = \alpha M_{p}$ in a radiation dominated universe. The modulus starts behaving as matter when the Hubble scale equals its mass. 
At this point, the energy density in the modulus is $\rho_{\rm mod} \sim m_{\rm mod}^{2} \phi_{\rm in}^2 =  m_{\rm mod}^{2} \alpha^2 M_{p}^2$ and
the energy density in radiation is $\rho_{\rm rad} \sim H^2 M_{p}^2 \sim m_{\rm mod}^2 M_{p}^2$. Thus, the ratio of the energy densities at the
time that the modulus start behaving like matter is given by $\alpha^{2}$
(so $\alpha<1$). The temperature when the modulus starts behaving like matter
is
\begin{align}\label{eq: TempMassRelMatter}
    T_{\rm mat} \sim \sqrt{m_{\rm mod} M_{p}}.
\end{align}
At any later time (with plasma temperature $T$), the ratio of energy densities is
\begin{align}
\label{eratio}
    {\Omega_{\rm mod}(T)  \over   \Omega_{\rm rad}(T)} \sim \alpha^2 \left(  {T_{\rm mat} \over T} \right) \sim 
    \alpha^{2} \left( {\sqrt{m_{\rm mod} M_{p}} \over T}\right) .
\end{align}

 Now, let us turn to the bounds. Recall that the CMB injection bounds arise from energy injection
(from decay of moduli particles) around the time of last scattering. In the context of moduli production via vacuum misalignment, the injection bound is therefore relevant only if
\begin{align}
\label{condition}
    T_{\rm mat} > T_{\rm CMB} \implies m_{\rm mod}\gtrsim 1\times 10^{-56} M_{p}\sim 3\times 10^{-29}\ {\rm eV},
\end{align}
where we have used the CMB temperature to be $\sim 0.3$
eV. Note that this condition is trivially met in the range of applicability of the injection bounds \eqref{massrange}.
The CMB injection bounds are obtained by making use of 
\eqref{eq: lifetime1} (as moduli produced by vacuum misalignment 
are non-relativistic) 
in \eqref{eq: injectionbound}. Note that \eqref{eq: injectionbound} involves the ratio of the energy density in dark matter
and the modulus. This can be expressed in terms of $\alpha$ and 
$m_{\rm mod}$ by making use of \eqref{eratio} and the observed value of
$\Omega_{\textrm{rad}} \large{/} \Omega_{\textrm{DM}}$ ($\sim 2.06\times 10^{-4}$), which is approximately constant from decoupling to today. We find the bound
\begin{eqnarray}
\label{injevac}
  m_{\rm mod}<\frac{2\times 10^{-27}}{\alpha^{4/7}}M_{p} \sim \frac{6}{\alpha^{4/7}}\ {\rm eV} .
\end{eqnarray}
This bound applies when the abundance obeys \eqref{fraccons} or 
\begin{equation}
\label{inifrac_misalign}
  \Omega_{\mathrm{mod}}\gtrsim 10^{-10}\Omega_{\rm DM} \implies m_{\rm mod} >
  \frac{2\times 10^{-75}}{\alpha^4}M_{\textrm{p}}\sim \frac{5\times 10^{-48}}{\alpha^4}\ \textrm{eV} ,
\end{equation}
using \eqref{eratio} and using the observed value of $\Omega_{\rm rad} \large{/} \Omega_{\rm DM}$.

In contrast, the over closure bound is obtained by simply demanding  that  $\Omega_{\rm mod}$ is less than the  observed fractional energy density of dark matter today. This similarly yields
\begin{align}
\label{overvac}
m_{\textrm{mod}}<\frac{2\times 10^{-55}}{\alpha^4}M_{\textrm{p}}\sim \frac{5\times 10^{-28}}{\alpha^4}\ \textrm{eV}.
\end{align}
The over closure bounds apply generally; specifically, the conditions \eqref{fraccons}, \eqref{massrange}  and \eqref{condition}  are not necessary.
However, for masses below $\sim 10^{-28}$ eV, the condition \eqref{overvac} is non trivial for $\alpha \gtrsim 1$. This would require not only super-Planckian initial displacements, but also an approximately quadratic form of the modulus scalar potential for several Planck distances, which is highly unnatural. This sets  a lower mass scale for the bound.

\begin{figure}[t]
    \centering
    \includegraphics[width=0.8\linewidth]{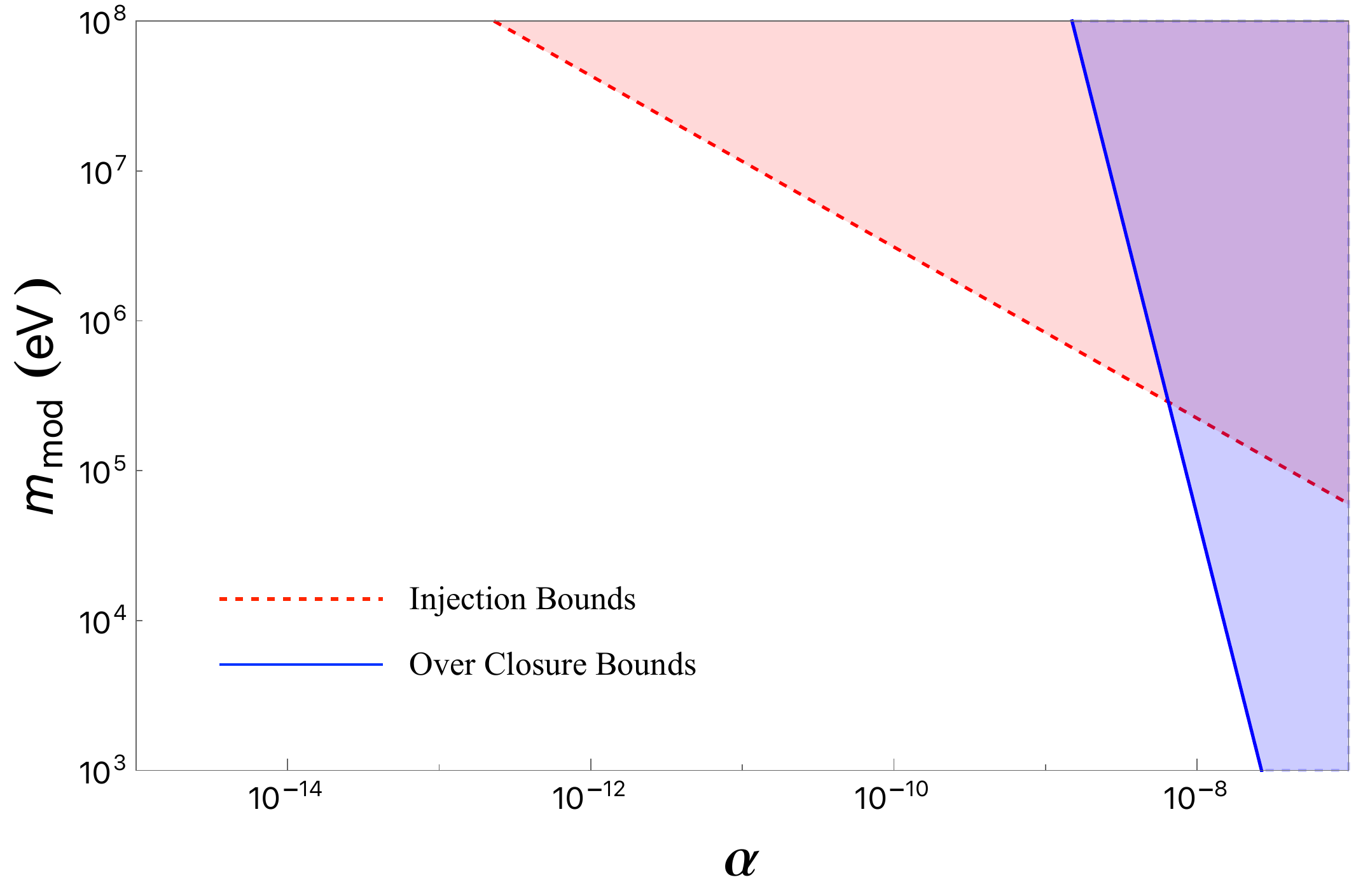}
    \caption{CMB injection (red) and overclosure (blue) bounds on moduli produced via vacuum misalignment with initial angular displacement $\alpha$. Excluded regions are shaded. The mass range is as per \eqref{mrangefinal}.}
    \label{fig:boundsinidis}
\end{figure}

We summarize our bounds for production by initial displacement in Figure \ref{fig:boundsinidis}.
Note that \eqref{injevac} and \eqref{overvac} scale with different powers of $\alpha$, so the dominant bound is determined by the value
of $\alpha$. The injection bound is stronger for $\alpha \lesssim 6\times10^{-9}$, i.e., intermediate  or smaller values of initial displacement $\lesssim 1\times 10^{10}$ GeV. 
We also see that we are justified in ignoring details of the injection bounds for masses below $10^3$ eV, since the parameter space that would be excluded by the injection bounds are already also excluded by the overclosure bound.

\subsection{Decays/Emissions}

Our discussion in  section \ref{sec:review} implies that the CMB injection bounds are relevant for
light moduli (see equation \eqref{eq:light}). When produced via decay/emissions, they are relativistic
(at the time of production).  We begin with a general analysis for such moduli; specific processes with details 
will be discussed later in the section.

   At the end of reheating, the typical momentum of such moduli is of the order of the 
   plasma temperature. Since their interactions are of gravitational strength, they do not thermalise --- they freestream  during the cosmological evolution. Initially, they behave  as radiation but start behaving as matter (i.e., become nonrelativistic) when the temperature
   of the universe becomes of the order of $m_{\rm mod}$. In analogy with our discussion in the previous subsection, we will refer to this temperature as $\hat{T}_{\rm mat}$. Furthermore, we  define
   $\hat{\alpha}^{2}$ as the ratio of the energy density in the modulus and
   the Standard Model plasma at this time:\footnote{The ratio is approximately the same at higher temperatures (the moduli and SM plasma both behave as radiation at higher temperatures).}
   \begin{equation}
   \label{eq:hatalphadef}
   \hat{\alpha}^{2} \equiv
     { \Omega_{\rm mod} (\hat{T}_{\rm mat}) \over  \Omega_{\rm rad}(\hat{T}_{\rm mat}) } .
   \end{equation}
If $\hat{T}_{\rm mat} (\approx m_{\rm mod})$ is less than the temperature of the big bang 
nucleosynthesis $(T_{\rm BBN})$, then $\hat{\alpha}^{2}$ is directly related to the contribution of the moduli to $\Delta N_{\rm eff}$. A single species of neutrino carries $7/43$ of the energy density of the Standard Model plasma,
thus $\Delta N_{\rm eff} \approx 43 \hat{\alpha}^{2}/7$ (neglecting the effects of change in $g_{*}$ between $T_{\rm BBN}$
and $\hat{T}_{\rm mat}$).  

As in the previous subsection, for plasma 
temperatures below $\hat{T}_{\rm mat}$, the relative ratio of the
energy density of the moduli and plasma is 
\begin{align}
\label{eratio2}
    {\Omega_{\rm mod}(T)  \over   \Omega_{\rm rad}(T)} \sim \hat{\alpha}^2 \left(  {\hat{T}_{\rm mat} \over T} \right) \sim 
    \hat{\alpha}^{2} \left( {{m_{\textrm{mod}} \over T}}\right) .
\end{align}
With this result, we can obtain the injection bounds. The condition \eqref{massrange} implies that the moduli
are non-relativistic at the time of decoupling, so the injection bounds \eqref{eq: injectionbound} apply.
The abundance \eqref{eratio2} along with the lifetime \eqref{eq: lifetime1} yield a bound 
\begin{equation}
\label{eq:Relhighmass}
m_{\textrm{mod}}<\frac{5\times10^{-24}}{\hat{\alpha}^{1/2}}M_{\textrm{p}}\sim\frac{10^4}{\hat{\alpha}^{1/2}}\textrm{eV}.
\end{equation}

Setting $T=T_0$, \eqref{fraccons} is
\begin{equation}
\label{inifrac_decays}
  \Omega_{\mathrm{mod}}\gtrsim 10^{-10}\Omega_{\rm DM} \implies m_{\rm mod} > \frac{5\times10^{-38}}{\hat{\alpha}^2}M_{\textrm{p}}
\sim\frac{1
\times 10^{-10}}{\hat{\alpha}^2}\textrm{eV} \
\end{equation}
Similarly, the over closure bound is
\begin{equation}
\label{relclosure}
m_{\textrm{mod}}<\frac{5\times10^{-28}}{\hat{\alpha}^2}M_{\textrm{p}}
\sim\frac{1}{\hat{\alpha}^2}\textrm{eV}
\end{equation}
For masses below the electron volt
scale, the condition \eqref{relclosure} is non trivial only for $\hat\alpha > 1$. This would correspond to $\Delta N_{\rm eff} > 1$, which is ruled
out by observations \cite{Planck:2018vyg}, so the $\Delta N_{\rm eff}$ bounds are more stringent in this mass range.

\begin{figure}[t]
    \centering
    \includegraphics[width=0.8\linewidth]{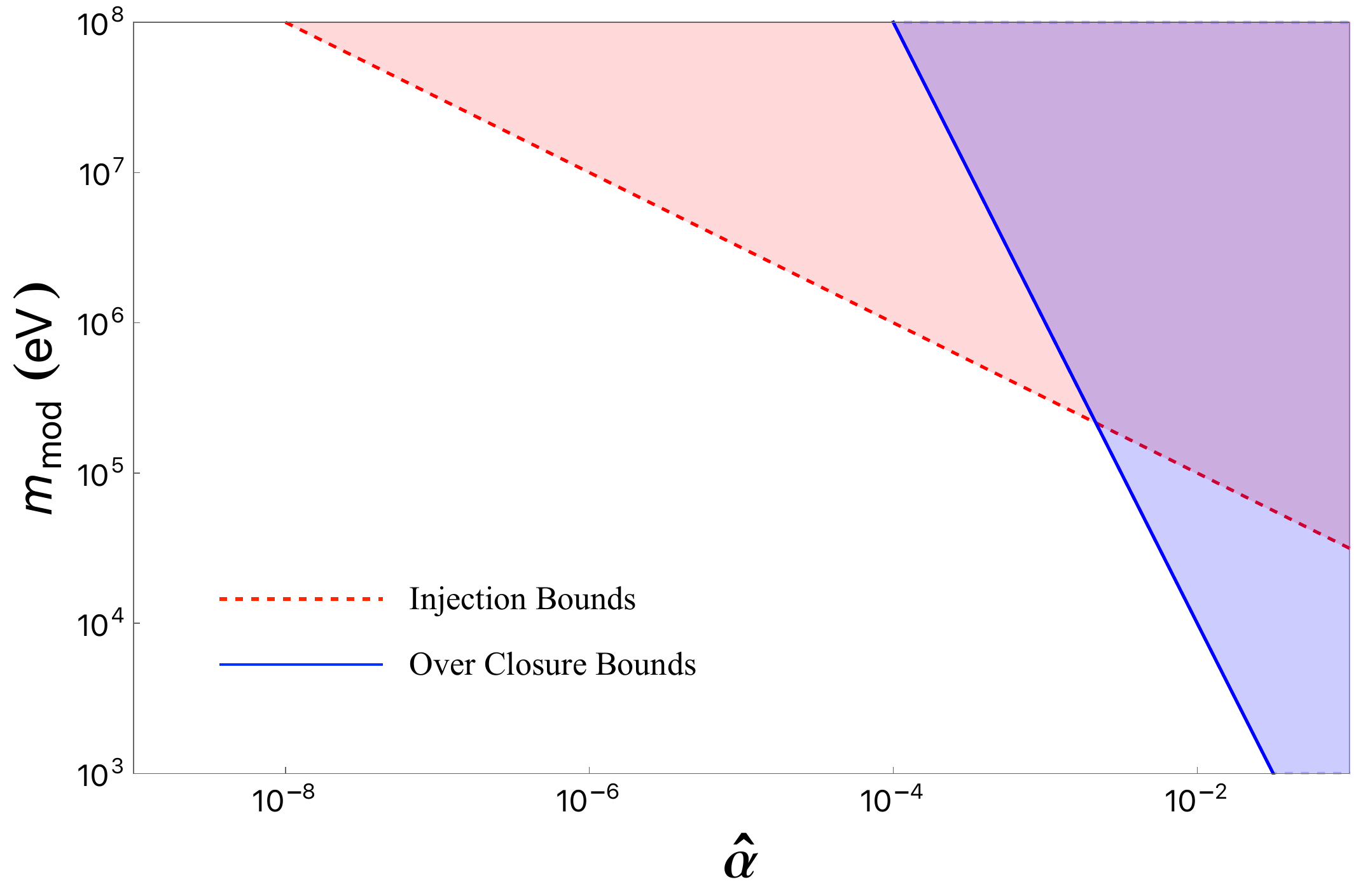}
    \caption{CMB injection (red) and overclosure (blue) bounds on moduli produced via decay or emission.
    $\hat{\alpha}$, defined in Eq.~\eqref{eq:hatalphadef}, compares the energy density in moduli with that of radiation when the moduli become nonrelativistic.
    In the misalignment mechanism, $\hat{\alpha}$ coincides with the initial angular displacement.
    The mass range is as per \eqref{mrangefinal}.}
    \label{fig:boundsdecay}
\end{figure}

As in the case of production via initial displacement, the overclosure bound   \eqref{relclosure} 
and injection bound \eqref{eq:Relhighmass} have different $\hat{\alpha}$ scaling. The injection bound dominates for
\begin{equation}
\label{highinject}
  \hat{\alpha}\lesssim 2\times 10^{-3}
\end{equation}
as shown in Figure \ref{fig:boundsdecay}.

     So far, the discussion has been general. Now,
let us analyze specific mechanisms for the production of relativistic moduli. The most obvious is
from the decay of the inflaton (or any heavy scalar dominating the energy density of the universe). In such 
a scenario, the fractional energy density is determined by the inflaton decay branching ratios. The parameter
$\hat{\alpha}$ (defined in \eqref{eq:hatalphadef}) is given by
   \begin{equation}
   \label{eq:hatalphadef2}
   \hat{\alpha}^{2} \equiv
     { \Omega_{\rm mod} (\hat{T}_{\rm mat}) \over  \Omega_{\rm rad}(\hat{T}_{\rm mat}) } = { B_{\rm mod} \over B_{\rm SM} },
   \end{equation}
where $B_{\rm mod}$ and $B_{\rm SM}$ are the inflaton decay branching ratios of the modulus and the SM. 
Therefore, Figure \ref{fig:boundsdecay} shows bounds on the modulus mass and the (square root) of the branching ratio.
In addition to this, there are two other mechanisms which are of interest: emission from the
 SM plasma and production from an early universe in a Hagedorn phase.

\subsubsection{Emission from  SM Plasma}\label{sec:sm-plasma}

We now discuss an irreducible production channel: freeze-in from the primordial plasma.
Very weakly coupled particles do not reach thermal equilibrium at any temperature, but the primordial plasma produces them in a suppressed manner, as a byproduct of $2 \to 2$ scattering involving three other particles that are in thermal equilibrium.
This process has been studied in the context of nonthermal production of photons from a quark-gluon plasma~\cite{Kapusta:1991qp,Arnold:2001ms}, and in the context of early universe physics for gravitini~\cite{Bolz:2000fu}, QCD axions~\cite{Braaten:1991dd,Graf:2010tv} and their  saxions~\cite{Graf:2012hb} in SUSY theories, right-handed neutrinos~\cite{Besak:2012qm} and gravitons~\cite{Ghiglieri:2015nfa,Ghiglieri:2020mhm,Ringwald:2020ist}.

For moduli, the logic is as follows.
The particle $\phi$ couples to the propagator of SM fields with gravitational strength, so there is a non-renormalisable, three-point vertex which can allow for the process $(SM,SM) \to (SM,\phi)$
(the specific coupling is a question for the UV completion, which we discuss at the end of the section).
At high temperature, where the masses of SM fields can be neglected, diagrams which involve soft gluon exchange yield IR divergences, which are then cutoff by plasma effects. The calculation of the relevant amplitude includes a hard-thermal-loop resummation~\cite{Braaten:1989mz,Braaten:1991dd} and yields a logarithmically enhanced production rate, proportional to $g_3^2\log(1/g_3)$, with $g_3$ the QCD coupling at the energy scale of interest
(we use $g_3$ here because gluon exchange dominates the process in the Standard Model).
We will consider couplings of the form
\begin{equation}
    \frac{\mathcal{L}}{\sqrt{-g}}=-\lambda\frac{\phi}{2M_p} G^a_{\mu \nu}G^{a,\mu \nu},
\end{equation}
with $G_{\mu\nu}$ the canonically normalised gluon field strength and $\lambda$ as the coupling constant. In what follows we focus on the QCD-mediated\footnote{Similar IR divergences due to intermediate fermionic states could appear, but this is not the case in the present calculation. Even if present they would not change our results qualitatively, since they would similarly be resummed giving an additive contribution of order $g_3^2 \log (1/g_3^2)$ times group theory factors.} production rate given by the logarithmic enhancement due to the IR divergence described above.
Concrete calculations, pioneered in~\cite{Braaten:1991dd}, involve separating the production rate in that mediated by soft bosons (with incoming momentum $q_s \ll k_c$) and hard bosons, with $k_c \ll q_h$, and the cutoff $k_c$ is set at an arbitrary scale satisfying $g_3 T \ll k_c \ll T$.
In practice the former calculation involves taking the imaginary part of the gluon-contribution to the particle's self energy, with one of the gluons having a hard-thermal-loop resummed propagator~\cite{Braaten:1989mz}.
The coefficient of the logarithmic divergence matches that of the hard part, which usually computes the rate from the 2-2 scattering amplitude.
For the particular case of axion and saxion production from a non-abelian plasma the production rate has been calculated in~\cite{Graf:2010tv,Graf:2012hb} to be, at leading-log order,\footnote{Notations match under the identification $\lambda/M_p=g_3^2/(16\pi^2f_{PQ})$.}
\begin{equation}\label{eq:production-rate}
    \frac{d\Gamma}{d^3k}\simeq\frac{3 \lambda^2 m_g^2 (N_c^2-1)T}{32\pi^4 M_p^2} \log\left( \frac{T^2}{m_g^2}\right)
    \frac{e^{-E/T}}{1-e^{-E/T}} \, .
\end{equation}
Here the thermal gluon mass is $m_g^2=g_3^2T^2(N_c+n_f/2)/9$, with $N_c$ the dimension of the fundamental representation of the gauge group in question (for QCD, $N_c=3$) and $n_f$ the number of fermions transforming under the representation $R$ of the group, weighted by the Dynkin index of the representation (so, for QCD, $n_f=12/2=6$).

Before proceeding to the calculation of the abundance, we note that Eq.~\eqref{eq:production-rate} also describes the production rate of axion-like particles with a coupling given by 
\begin{equation}\label{eq:axion-coupling}
    \frac{\mathcal{L}}{\sqrt{-g}}=-\frac{\lambda}{4} \frac{\phi}{M_p }
    \epsilon^{\mu\nu\rho\sigma} 
    G_{\mu\nu}^a G_{\rho \sigma}^a \, .
\end{equation}
Thus, similarly to the previous section, our bounds also apply to parity-odd scalars provided their coupling is of Planckian strength.

An evolution equation for the total number of moduli particles is thus given by
\begin{equation}\label{eq:production-sm}
    \frac{\partial n}{\partial t}+3Hn=
    \frac{ 3\lambda^2  (N_c^2-1)m_g^2T^4}{8\pi^3 M_p^2} \log\left( \frac{9}{g_3^2(N_c+n_f/2)}\right)
    \int_0^\infty{dx} \, \frac{x^2}{e^{x}-1} \equiv A_1 \frac{T^6}{M_p^2} \, ,
\end{equation}
where
\begin{equation}
    A_1=\frac{\lambda^2 \zeta(3)g_3^2(N_c+n_f/2) (N_c^2-1)}{12\pi^3}\log\left( \frac{9}{g_3^2(N_c+n_f/2)}\right) 
    \, 
\end{equation}
is a constant, modulo running of couplings.
In the case of QCD at $10^{15}$ GeV we have
\begin{equation}
    A_1 \simeq 3 \cdot 10^{-2} 
    \left(\frac{\lambda}{1/\sqrt{2}}\right)^2
    \left(\frac{g_3}{0.5}\right)^2\frac{N_c+n_f/2}{3+3}\frac{N_c^2-1}{8}\frac{\log\left(\frac{9}{g_3^2(N_c+n_f/2)}\right)}{\log(6)} \, .
\end{equation}
The constant $\lambda$ is model-dependent and order one in case of direct coupling between the modulus and the gluons.\footnote{We have chosen the fiducial value ($\lambda = {1 \over \sqrt 2}$) motivated by the string theory --  the dilaton's coupling to
gauge degrees of freedom on a D3 brane. See the end of the subsection for further discussion.}
The temperature dependence of the production rate in Eq.~\eqref{eq:production-sm} is interpreted as the background ($\rho_{bck}\sim T^4$) sourcing a fraction of its energy density into moduli through a process penalised by a gravitational-strength coupling $(T/M_p)^2$.
The non-renormalizability of the interaction thus implies that the largest contribution to the particle number density is sourced at the highest energies reached by the plasma, namely the reheating temperature, $T_{\rm reh}$.\footnote{This observation has led to the proposal that the amplitude of the Cosmic Gravitational Wave Background is a ``thermometer" of the reheating temperature~\cite{Ringwald:2020ist}.}
More concretely, we can integrate Eq.~\eqref{eq:production-sm} from the reheating time, $t_{\rm reh}$, to a fiducial time $t_*$ to find (neglecting pre-existing moduli):
\begin{equation}
    n(t_*)= \frac{T_{\rm reh}}{M_p}\frac{3A_1}{\pi }\sqrt{\frac{10}{g_*(t_{\rm reh})}}T_*^3 \frac{g_{*}(t_*)}{g_{*}(t_{\rm reh})} \, .
\end{equation}
where $g_{\star}(t)$ denotes the effective number of relativistic degrees of freedom at time $t$. In deriving the number density of moduli, we have used that the universe is in a radiation domination between reheating and time $t_\star$ (which is valid for the mass range of moduli we consider). We conclude that the number density at any time is directly proportional to the reheating temperature.

\begin{figure}[t]
    \centering
    \begin{subfigure}[t]{0.47\textwidth}
        \includegraphics[width=\textwidth]{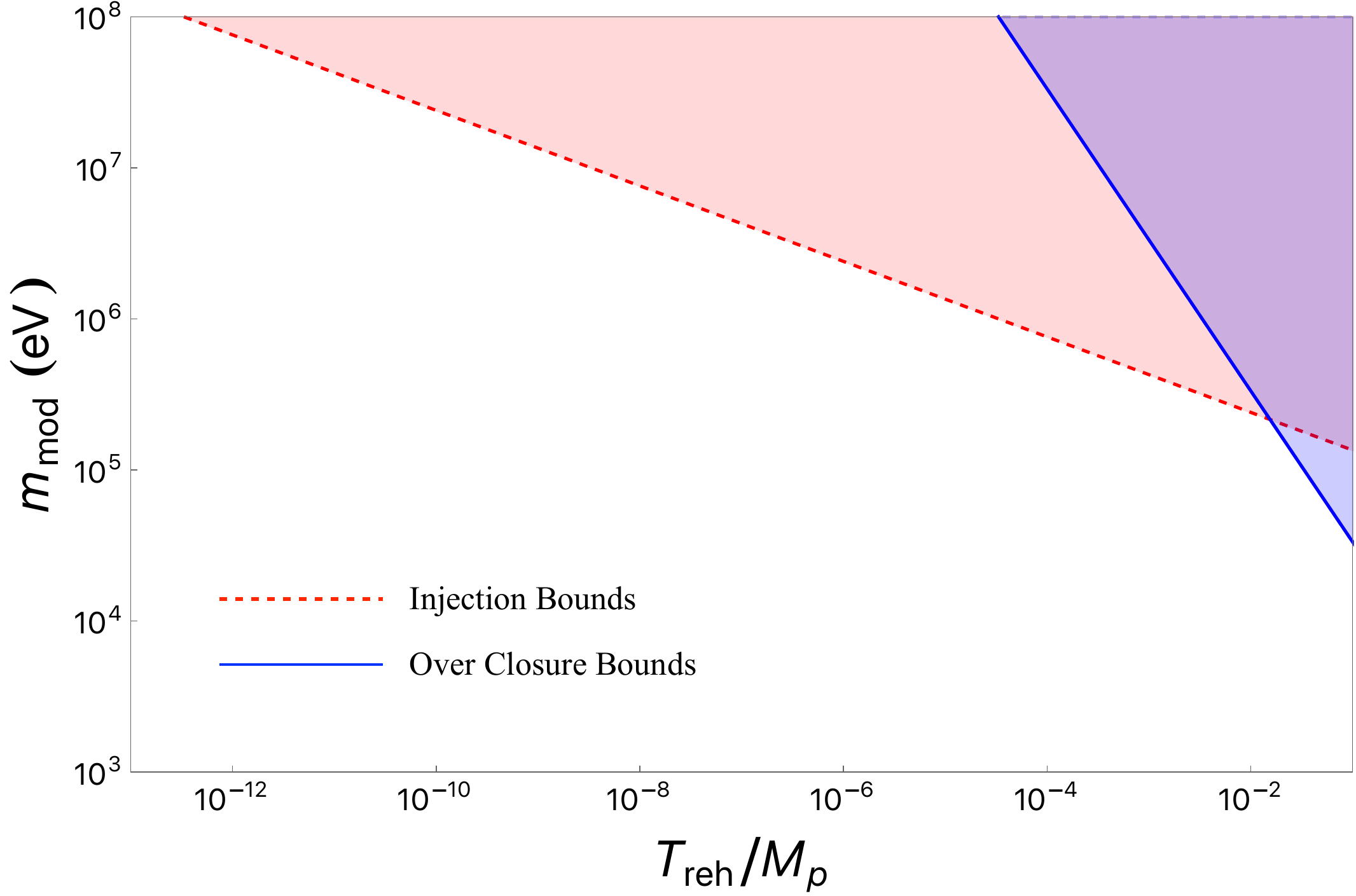}
        \caption{Standard Model production}
        \label{subfig:SMcase}
    \end{subfigure}
    \begin{subfigure}[t]{0.47\textwidth}
        \includegraphics[width=\textwidth]{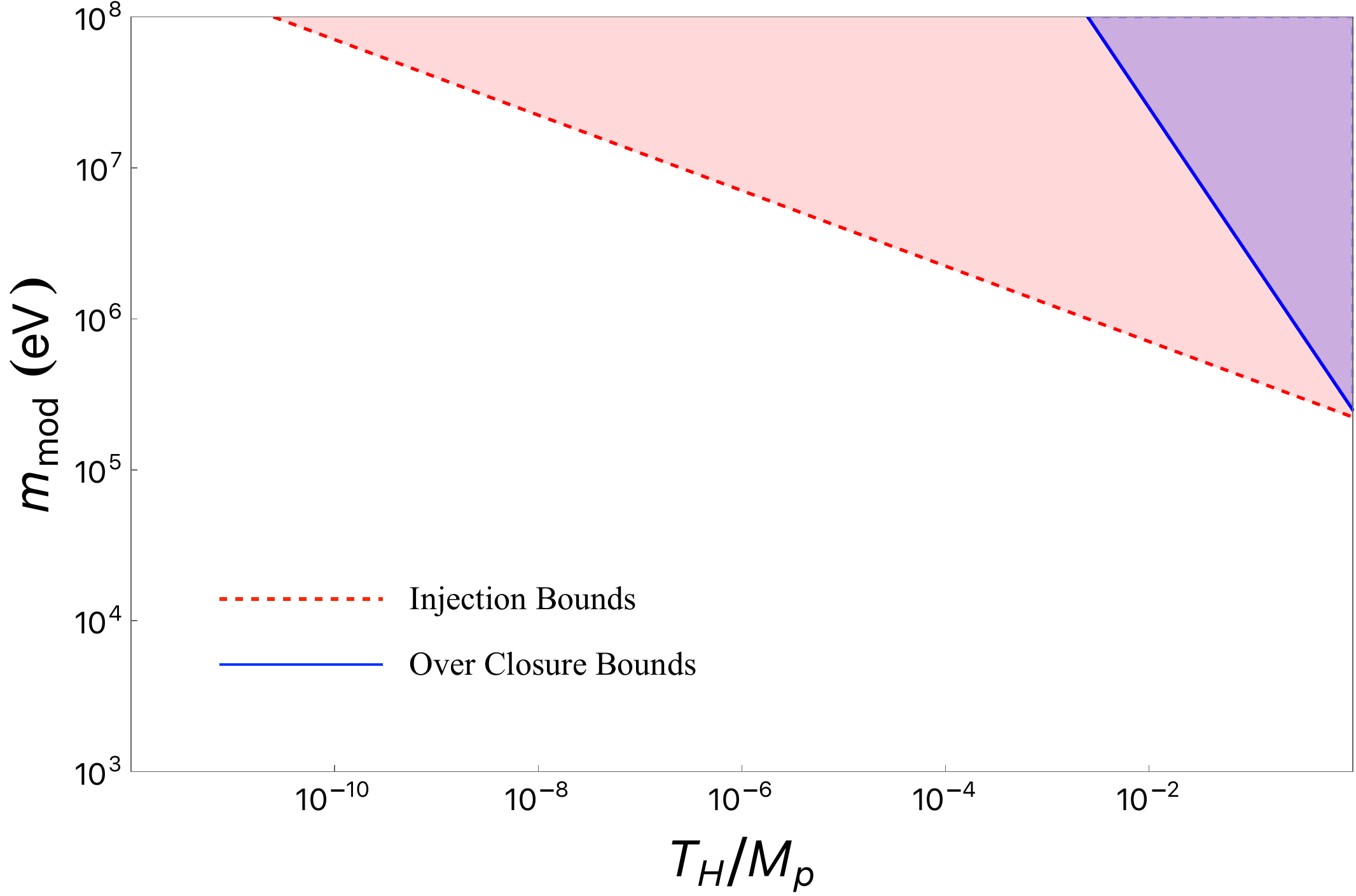}
        \caption{Hagedorn Phase production}
        \label{subfig:Hagedorncase}
    \end{subfigure}
    \caption{CMB injection (red) and overclosure (blue) bounds on moduli produced via freeze-in in the Standard Model (left) or a Hagedorn phase (right).
    In these mechanisms the reheating (Hagedorn) temperature $T_{\rm reh}$ ($T_H$) sets the moduli population, 
    which allows to bound $m_{\rm mod}$ as function of $T_{\rm reh}$ ($T_H$).
   We use $A_1=3\times 10^{-2}$ for SM production and $M_s^4=\rho_{\textrm{end}}$ for Hagedorn production in these plots. }
    \label{fig:boundsdecaySMandHagedorn}
\end{figure}

Lastly, we can evaluate the number density at a time $t_{\textrm{mat}}$ when $\hat{T}_{\textrm{mat}}=m_{\textrm{mod}}$ and use it to calculate $\hat{\alpha}$ (cf. Eq.~\eqref{eq:hatalphadef}) as
\begin{equation}
    \hat{\alpha}^2=\frac{m_{\textrm{mod}} n(t_{\textrm{mat}})}{\frac{\pi^2}{30}g_*(t_{\textrm{mat}})\hat{T}^4_{\textrm{mat}}}=
    \frac{T_{\rm reh}}{M_p}\frac{90A_1}{\pi^3 }\frac{\sqrt{10}}{(g_*(t_{\rm reh}))^{3/2}} \, .
\end{equation}
Taking $g_*(t_{\rm reh})\simeq 107$ (SM effective degrees of freedom at high temperature), we have as a fiducial value for $\hat{\alpha}^2$ as
\begin{equation}\label{eq: SMalpha}
    \hat{\alpha}^2 \simeq  3 \cdot 10^{-4}  \left(\frac{A_1}{3 \cdot 10^{-2}}\right)\frac{T_{\rm reh}}{M_p}\, .
\end{equation}
Because the number density is a function of the reheating temperature, we obtain bounds on moduli masses as a function of $T_{\rm reh}$.
This is illustrated in Fig.~\ref{subfig:SMcase}, where we insert Eq.~\eqref{eq: SMalpha} into Eqs.~\eqref{inifrac_decays} and~\eqref{relclosure} to apply the bounds directly to $T_{\rm reh}$.

We close this section with two remarks.
First, in string theory what the specific coupling of the particle in question to SM fields is depends on the UV realisation of the SM, but the existence of a gravitational coupling from QCD to moduli fields is model-independent.
Consider the SM embedded in a stack of branes and $\phi$ the overall volume modulus. 
$\phi$ couples to the trace of the stress tensor, $T_M^M$.
For the gauge fields of a D3-brane this is zero off-shell, so there is no tree-level coupling between these fields.
But there is a direct coupling to the string dilaton $S$
(indeed the gauge kinetic function in this case is $f_{D3}=S$ at tree level).
If QCD is realised in D7 branes there is instead a direct coupling to the size of the 4-cycle the branes wrap at the level of the kinetic term.
These examples are manifestation of a general principle: in string theory there are no free parameters and gauge couplings are given by expectation values of moduli fields, so there is always a production rate provided the reheating temperature exceeds the mass of the field.

The second point to discuss is the relation between the phenomenology of these moduli and the CGMB.
The same production mechanism (with gravitons rather than moduli as external legs) leads to a stochastic gravitational wave background whose peak frequency today lies at CMB frequencies~\cite{Ghiglieri:2015nfa,Ghiglieri:2020mhm,Ringwald:2020ist}.
The amplitude of the spectrum grows linearly with the reheating temperature, similar to the production rates in this section.
While the CGMB signal, even at reheating temperatures close to the Planck scale, lies beyond reach of any foreseeable detector, it is an interesting fact that the non-renormalisability of gravity provides a way to (in principle) probe the reheating temperature from above.
Our bounds are similarly stronger for large reheating temperatures.

\subsubsection{Hagedorn Phase Production}\label{sec: hagedornproduction}
In this subsection, we will consider moduli production from a hot stringy phase of the early universe called the Hagedorn phase. The  conditions under which the universe can remain in equilibrium in such a phase have been outlined in \cite{Frey:2005jk, Frey:2023khe, Frey:2024jqy, Chakraborty:2026hob}. 
To make a specific estimate, we consider volume modulus production by a thermal open string state on D3-branes with warped extra dimensions, although  the moduli production rate is model-dependent. 
A detailed review of the Hagedorn phase in this class of backgrounds can be found in appendix \ref{sec: generalcal-app}.

The equilibrium energy density of the open and closed strings in the Hagedorn string phase are \cite{Frey:2024jqy}
\begin{align}\label{eq: stringbackEdensity}
\rho_o = M_s^2\int_{l_c}^{\infty} dl' l' \tilde{n}_o(l')\simeq (N_DL M_s)^2  M_s^4, \quad \rho_c \simeq M_s^4
\end{align}
up to leading order in $l_c/L$ where $l_c$ is a cut-off for the length of the strings in the distribution, $N_D$ is the number of D-branes,  $\tilde{n}_o(l)$ denotes the number of open strings with length between $l$ and $l+dl$, and $M_s$ is the local string scale. At equilibrium, open strings dominate the energy density under the condition $N_D M_sL\gg 1$ in our setup. 
In the estimates \eqref{eq: stringbackEdensity}, we have ignored a common phase space factor in the energy densities $\rho_o,\rho_c$.\footnote{Following \cite{Deo:1988jj,Deo:1989bv}, this prefactor for three noncompact dimensions is approximately $10^{-4}$.} 
To account for that factor as well as the fact that the transition from an open string Hagedorn phase to radiation is continuous (rather than a sharp phase transition), we will take $\rho_{\mathrm{end}}$, the energy density at the end of the Hagedorn phase, to be a free parameter of order $M_s^4$.

The coupling of gravitons to long open strings in the Hagedorn gas can be analysed in a model independent way \cite{Frey:2024jqy}. On the other hand, the couplings of moduli depends on the modulus in question and are model dependent. We will focus on the volume modulus in type IIB string theory compactified over a Calabi-Yau orientifold with self-dual flux. 
The volume modulus can be taken to be stabilized via either KKLT \cite{Kachru:2003aw} or a LVS \cite{Balasubramanian:2005zx} type scenario giving the volume modulus mass $m_{\textrm{mod}}$. In this case, we have carried out an explicit computation of the coupling to long strings in appendix \ref{sec: generalcal-app}. The final result is that the coupling is the same as couplings to gravitons \cite{Frey:2024jqy} up to a factor of $1/\sqrt{6}$ when the strings are localized in a sufficiently warped region. 
Using this we can write the decay rate of the long string by volume modulus emission as
\begin{align}\label{eq: decayrate}
    \frac{d\Gamma_{o,\mathfrak{c}}}{d\omega}=\frac{\tilde{A}}{6}\left(\frac{M_s}{M_{p}}\right)^2 l M_s (\omega/T_H)^3 \frac{e^{-\omega/T_H}}{(1-e^{-\omega/2T_H})^2} ,
\end{align}
where $\tilde{A}=(T_H/M_s)^3/4\pi^3$, $T_H$ is the Hagedorn temperature, $l M_s$ is the length of the string in string units, and $\omega$ is the frequency of the emitted modulus. This is the modulus version of equation (3.21) in \cite{Frey:2024jqy} apart from an extra factor of 1/6, which is the relative string/modulus coupling compared to the string/graviton coupling.

Similar to graviton production from hot open string plasmas in \cite{Frey:2024jqy}, radiated moduli are relativistic, since the mass range we consider is $m_{\textrm{mod}}\ll T_H$. 
Also, the volume modulus is produced out of equilibrium; simply speaking, not enough moduli are produced to reach an equilibrium value because the string/modulus coupling is suppressed by $M_s/M_p$. This coupling is suppressed because the profile of the volume modulus (even when it is stabilized) is spread through the extra dimensions rather than localized to the warped throat where the long strings are localized \cite{Agarwal:2025rqd}. Now, the total moduli production rate is the single string emission rate integrated over the open string distribution and the modulus frequency. Since the modulus emission rate (\ref{eq: decayrate}) is proportional to the string length, the rate of moduli production is \comments{($\Omega_\perp\sim\Omega_\parallel\sim 1$)}
\begin{align}
    \frac{\partial n}{\partial t}+3Hn =\kappa\,\frac{\tilde{A}}{6}\left(\frac{M_s}{M_{p}}\right)^2 \left(\frac{T_H}{M_s}\right) \rho_o(t) ,
\end{align}
where $\kappa=(96\zeta(3)-16\pi^4/15)$ arises from the frequency integral and $n(t)$ is the modulus number density. Because the Hagedorn strings redshift like matter, this can be integrated to give 
\begin{align}
    n(t)=\frac{\kappa\tilde{A}}{3\sqrt{3}}\left(\frac{T_H}{M_s}\right) \rho_{\mathrm{end}} \left(\frac{M_s}{M_{p}}\right)^2\frac{M_{p}}{\sqrt{\rho_{\rm{end}}}}\left(\frac{a(t_{\rm{end}})}{a(t)}\right)^3 ,
\end{align}
where the \enquote{end} subscript denotes values of the respective quantities at the end of the Hagedorn phase. 
Since the entropy density of long open strings is $s=\rho_o/T_H$, adiabatic expansion to the temperature $\hat{T}_{\mathrm{mat}}$ where the modulus begins behaving like matter leads to 
\begin{align}\label{eq: Hagedornalpha}
\hat{\alpha}^2=\frac{\Omega_{\textrm{mod}}(\hat{T}_{\textrm{mat}})}{\Omega_{\textrm{rad}}(\hat{T}_{\textrm{mat}})}=\frac{4\kappa\tilde{A}}{9\sqrt{3}}\left(\frac{m_{\text{mod}}}{\hat{T}_{\textrm{mat}}}\right)\left(\frac{M_s}{M_p}\right)\frac{T_H^2}{\sqrt{\rho_{\mathrm{end}}}}\simeq 
4\times 10^{-6}\left(\frac{M_s^4}{\rho_{\mathrm{end}}}\right)^{1/2} \left(\frac{T_H}{M_p}\right).
 \end{align}
where we have used $M_s=2\pi\sqrt{2}\,T_H$ and assumed that the energy of the open string bath becomes Standard Model radiation.\footnote{Our computation of $\hat{\alpha}^2$
is based on the simple model in \cite{Frey:2024jqy}, 
where one does not have a complete realization of the
Standard Model degrees of freedom. A setting with a
complete realization of the Standard Model degrees of 
freedom and a non-standard  cosmological evolution of the Standard Model degrees of freedom can have a significantly larger value of $\hat{\alpha}^2$. See 
\cite{Frey:2024jqy} for a more detailed discussion of
such effects.} We have assumed that the effective number of degrees of freedom for entropy and energy are the same at $\hat T_{\mathrm{mat}}$.
 
Figure \ref{subfig:Hagedorncase} translates the bound \eqref{eq:Relhighmass} into a bound on the Hagedorn temperature due to volume modulus production for fiducial value $\rho_{\mathrm{end}}=M_s^4$. 
The injection bound dominates the overclosure bound across the entire mass range considered because \eqref{eq: Hagedornalpha} implies $ \hat{\alpha}\ll 10^{-6}$ for $T_H<M_p$ and the injection bound dominates when $\hat{\alpha} \lesssim 2\times 10^{-3}$ (see \eqref{highinject}). However, it is important to note that
a small value of $\rho_{\mathrm{end}}$ (related to phase space factors in the string number distributions) can increase $\hat\alpha$.

\section{Conclusions}\label{sec: Conclusion}

In this paper we have systematically evaluated CMB energy injection bounds on long-lived, light moduli fields ($10^{3} \ \text{eV}  \lesssim  m_{\rm mod} \lesssim 10^{8} \ \text{eV}$). Their gravitational strength couplings cause them to decay slowly and allow them to inject energy to the primordial Standard Model plasma at or near the epoch of CMB decoupling (for X-ray backgrounds bounds on moduli in this mass range see \cite{Kawasaki:1997ah}).
By examining the primary cosmological production channels (vacuum misalignment, heavy scalar/inflaton decay, thermal emission from the primordial plasma, and production from a Hagedorn phase of hot strings),  we have derived constraints that complement existing bounds in the literature.

   Let us contrast our study with other bounds on moduli:
   \begin{itemize}
  \item Current fifth force bounds require the mass of a scalar which mediates a force of gravitational strength between matter to be above $10^{-2}\ \text{eV}$ (see e.g. \cite{Adelberger:2003zx,Kapner:2006si}).
  Injection bounds instead apply in the unconstrained region $(10^{3} \ \text{eV}  \lesssim  m_{\rm mod} \lesssim 10^{8} \ \text{eV})$ as in Eq.~\eqref{mrangefinal}. 
  
  \item Overclosure considerations apply for
  $m_{\rm mod} \gtrsim 10^{-28}  \ \text{eV}$ for production by initial displacement and $m_{\rm mod} \gtrsim 1 \ \text{eV}$ in the context of thermal emission (see discussions after equations \eqref{overvac} and \eqref{relclosure}).
  We have seen in Figs.~\ref{fig:boundsinidis} and~\ref{fig:boundsdecay} that injection bounds are more constraining in a regime of parameter space.

  \item The cosmological moduli problem implies that $m_{\rm mod} \lesssim 30\ \text{TeV}$ is in tension with nucleosynthesis, assuming an order one initial misalignment.
  Under this assumption, the constraint extends to $m_{\rm mod} \gtrsim 10^{-28}$ eV
  for cosmology to be consistent with the measured evolution thereafter, since a modulus should not contribute a meaningful fraction to the energy density of the universe until matter-radiation equality (see Eq.~\eqref{overvac}).
  Thus, for order one initial misalignment, these considerations rule out the parameter space we have discussed.
  Our discussion shows that, even in the case where the initial misalignment is negligible (avoiding a CMP and related issues), the CMB has constraining power in a range of masses.

  \item CMB spectral distortions can be induced due to modulus decay into photons at times earlier than recombination, and even earlier decays are constrained by nucleosynthesis.
  This has been studied in~\cite{Balazs:2022tjl} in the context of axion-like particles produced via freeze-in, similar to our Section~\ref{sec:sm-plasma}.
  These complementary bounds should provide constraints at higher masses.
  
  \item Stellar cooling and other astrophysical bounds (see Ref.~\cite{Arza:2026rsl} for a recent review) apply for weakly coupled particles in the mass range of interest.
  In our case, however, they are not competitive because they can only probe couplings of the form of Eq.~\eqref{eq:axion-coupling} with magnitude of order $\lambda/M_p\sim (10^{10}\text{ GeV})^{-1} \gg 1/M_p$.
\end{itemize}

Our bounds are therefore complementary to other bounds in the literature and apply to any gravitationally coupled scalar field.
Future observational programs offer significant potential to refine these bounds. Next generation CMB polarization and anisotropy experiments, including the Simons Observatory \cite{SimonsObservatory:2025wwn}, Lite-Bird \cite{LiteBIRD:2024dbi}, PICO \cite{NASAPICO:2019thw} and CMB-S4 \cite{CMB-S4:2016ple} can drastically improve sensitivity to extra degrees of freedom and energy injection thresholds.

 Moreover, complementary observational windows will sharpen these constraints, including 21cm cosmology ~\cite{DAmico:2018sxd,Mitridate:2018iag,Sun:2023acy,Zhao:2025ddy}, CMB spectral distortions ~\cite{Zeldovich:1969ff,Chluba:2011hw}, $X$-ray and $\gamma$-ray astronomy ~\cite{Essig:2013goa,Boyarsky:2006zi}.
 Furthermore, if moduli decay preferentially into ultra-light axions or dark radiation in general, the constraints shift from electromagnetic energy injection to bounds on $\Delta N_{\rm{eff}}$.\footnote{For completeness, we have considered the decays of moduli to axions in one LVS model considered in \cite{Agarwal:2026mtm}, which has one ultralight axion. There is a gravitational-strength coupling through the axion kinetic term, but a potentially larger coupling exists in the potential. The modulus-axion-axion coupling in the potential is proportional to the axion mass squared, and decays of the modulus to the axion via this coupling are subdominant to those by gravitational-strength dimension 5 couplings in the region of moduli space where the modulus is light enough to be cosmologically long-lived. Of course, these conclusions may depend on the ultraviolet completion.}
 
Light gravitationally coupled particles provide a unique window into high energy and  early universe physics. As observational sensitivity advances, these constraints will continue to narrow the allowed parameter space for string compactifications and general light scalar field phenomenology. This will  better shape our understanding of the early universe.

\acknowledgments

We would like to thank Aaron Pierce and Evan McDonough for discussions. 
NA and ARF are supported by the Natural Sciences and Engineering Research Council of Canada (NSERC) via Subatomic Physics Discovery Grant 2026-00042.
AM would like to thank the Leinweber~Institute for Theoretical Physics for supporting his sabbatical visit at the University
of Michigan, Ann Arbor. 
RM is supported by the National Natural Science Foundation of China (NSFC) under Grant No. 12247103. FQ's research is funded by Tamkeen under the research grant to NYUAD ADHPG-AD457.
The research of GV is funded by a Research Fellowship from Gonville and Caius College.

\appendix
\section{Hagedorn Phase and Moduli Coupling to Long Strings}\label{sec: generalcal-app}

As mentioned in section \ref{sec: hagedornproduction}, we wish to consider the coupling of a modulus field to a thermal open string state on D3 branes in three non-compact spatial dimensions (plus time). Since the profile of the modulus in the extra dimensions depends both on the type of compactification and the particular modulus in the compactification, we consider the volume modulus in Calabi-Yau orientifold compactifications of the type IIB superstring with imaginary self-dual 3-form flux \cite{Dasgupta:1999ss,Greene:2000gh,Giddings:2001yu}\footnote{See \cite{Becker:1996gj} for early work on related M theory compactifications.} for specificity but stress that these results are likely model-dependent.

\paragraph{Thermal strings in warping:}
In warped compactifications, long strings are confined at the bottom of a warped throat in a string scale region, as argued by Jackson-Jones-Polchinski (JJP) \cite{Jackson:2004zg}. We call this region the JJP box; \cite{Frey:2005jk,Frey:2024jqy} extended the JJP scenario to string thermodynamics, and we largely follow their discussion here. 

The basic JJP argument is that fundamental strings experience a potential due to the warp factor, and their quantum (or thermal) fluctuation around the minimum is set by the curvature scale of the warp factor, which is expected to be slightly more than the 10D string length.
In comparison, for a gas of hot strings with typical length $L$, the typical extent of a string in a noncompact dimension is $L_{\rm rms}\sim\sqrt{L/M_s}$. 
Since $L_{\mathrm{rms}}$ is much larger than the size of the JJP box, the long thermal strings fill the JJP box but do not probe the rest of the extra dimensions; this is similar to how long strings fill a small torus.
Similarly, the intersection (and self-intersection) rate of long strings in the JJP box is independent of the string length $l$; this is also characteristic of a string gas on a torus smaller than $L_{\mathrm{rms}}$.
Therefore, we can treat thermal strings in warping as if they were on a roughly string-scale torus. This picture is also borne out by worldsheet calculations \cite{PandoZayas:2003jr,Canneti:2024iyn}.

For a gas of strings with typical length $L$ in a JJP box at the tip of a warped throat with $N_D$ D3-branes, 
the equilibrium energy density of the open and closed strings in the Hagedorn string phase are
\begin{align}
\rho_o = M_s^2\int_{l_c}^{\infty} dl' l' \tilde{n}_o(l')\simeq \frac{(N_DL M_s)^2}{\Omega_{\perp}} M_s^4, \quad \rho_c \simeq M_s^4
\end{align}
up to leading order in $l_c/L$, where $l_c$ is a cutoff length below which strings cannot be considered to be long (and are assumed to contribute negligibly).
The compact volume of the JJP box perpendicular to the D3-branes, as measured in string units, is $\Omega_\perp\gtrsim 1$ (for higher-dimension branes, there is a corresponding compact volume $\Omega_\|$ along the brane worldvolume). At equilibrium, open strings dominate the energy density if $N_D M_sL\gg \sqrt{\Omega_{\perp}}$ which is generically the case in the JJP box. Note that $1/L=M_s^2(\beta-\beta_H)$ gives the temperature of the gas.

For a well-controlled description of Hagedorn physics in cosmology, there are two necessary conditions. First, the cosmological expansion should be described in 4D effective field theory; second, the string gas should interact quickly enough to remain in equilibrium.
Considering the first, effective field theory should be valid when the Hubble scale is less than the KK scale $H\ll M_{KK}$, or $(N_D L M_s)/\sqrt{\Omega_\perp} (M_s/M_p)\ll M_{KK}/M_s$. This condition is also parametrically true if we demand that the energy density in an earlier inflationary stage is greater than the minimal density of a Hagedorn phase and also satisfies $H\ll M_{KK}$. As argued in \cite{Frey:2024jqy}, this condition can be satisfied by strings in either large volume compactifications or strongly warped regions (with a JJP box).
Regarding the second condition, the string gas remains in equilibrium if the equilibration rate is greater than or equal to Hubble scale. 
The equilibration rates follow from Boltzmann equations, given in general cases in \cite{Frey:2023khe} (see \cite{Lowe:1994nm,Lee:1997iz,Copeland:1998na} for earlier work on string Boltzmann equations); \cite{Frey:2024jqy} argued that the flat spacetime equilibration rates apply to cosmologies that are well-described by 4D effective field theory.
In that case, the ratio of the equilibration rate to the Hubble parameter is  $\Gamma/H\propto M_p/M_s$ for the equilibration of open strings with each other and open strings with closed strings, so the string interactions are fast enough to maintain local thermal equilibrium.

\paragraph{Coupling to long strings:}
Next, we can discuss the coupling of the volume modulus to long strings localized in a warped compactification (as in a JJP box). As mentioned, it is not possible to understand the couplings of moduli in a model independent way in contrast to the graviton. 
Therefore, we consider the volume modulus in Calabi-Yau orientifold compactifications of the type IIB superstring with imaginary self-dual 3-form flux \cite{Dasgupta:1999ss,Greene:2000gh,Giddings:2001yu}.
In these compactifications, the background metric is  
\begin{align}
    ds_0^2=e^{2\Omega_0}e^{2A_0(y)}\hat{\eta}_{\mu\nu}dx^{\mu}dx^{\nu} + e^{-2A_0(y)}\tilde{g}_{0,mn}dy^m dy^n.
\end{align}
where $e^{2A}$ is the warp factor,
\begin{align}
     e^{-2\Omega}=\frac{\int d^6y \sqrt{\tilde{g}}\,e^{-4A}}{\int d^6y \sqrt{\tilde{g}_0}}
\end{align}
is the Weyl factor required to convert to Einstein frame in 4D, and $\tilde{g}_{0,mn}$ is a fixed Ricci-flat Calabi-Yau metric. The subscript $0$ denotes the background value of the various fields in the metric. Among other supergravity backgrounds, the dilaton takes a fixed constant value.

This type of compactification has a (massless) volume modulus at tree level, which can be stabilized via nonperturbative effects (gaugino condensation, for example) as in \cite{Kachru:2003aw} or large volume scenarios \cite{Balasubramanian:2005zx}.
These effects act as a source for the bulk supergravity fields, perturbing them from the background described above; \cite{Agarwal:2025rqd} found that the profile of the stabilized volume modulus is therefore a small perturbation of its profile in the tree-level compactification, which we therefore use here.
The volume modulus fluctuation takes the form \cite{Giddings:2005ff,Frey:2008xw}
\begin{align}
    ds^2=e^{2\Omega(x)}e^{2A(x,y)}\hat{\eta}_{\mu\nu}dx^{\mu}dx^{\nu}+2e^{2\Omega_0}e^{2A_0(y)}\partial_\mu c(x) B_m (y)dx^{\mu}dy^m + e^{-2A(x,y)}\tilde{g}_{mn}dy^m dy^n
\end{align}
with spacetime-dependent warp factor
\begin{align}
    e^{-4A(x,y)}=e^{-4A_0(y)}+c(x).
\end{align}
Here $B_m(y)$ is an off-diagonal compensator which is sourced by the background warp factor and Weyl factor.

To find the coupling of the long string to moduli, we consider a quantum mechanical treatment (outlined in \cite{Frey:2024jqy} for gravitons, following the treatment of matter-light interactions in \cite{weinberg2015lectures}). The action in 10D Einstein frame is
\begin{align}
    S=\int dt \left(\frac{M_{10}^8}{2}\int d^3x\,d^6y \sqrt{-g_{10}} R_{10}+\cdots -\int d\sigma\ T_1\sqrt{-\gamma}\right),
\end{align}
where $M_{10}$ is the ten dimensional Planck mass, $T_1=e^{\frac{\Phi}{2}}/2\pi\alpha'$ is the tension of the long string in Einstein frame (with $\Phi$ the  dilaton), and $\gamma_{\bar{A}\bar{B}}$ is the pullback metric on the string worldsheet (bar denotes worldsheet coordinates $(\sigma,\tau=t)$).

To find the interaction with the modulus, we expand the long string action with respect to the metric fluctuation representing the modulus. To do this, we note that the pullback on the worldsheet can be written as
\begin{align}
    \gamma_{\bar{A}\bar{B}}=g_{AB}\partial_{\bar{A}}X^A \partial_{\bar{B}}X^B = \gamma_{0,\bar{A}\bar{B}} + \gamma_{1,\bar{A}\bar{B}}+\cdots
\end{align} 
where $\gamma_{0,\bar{A}\bar{B}}$ denotes the pullback from the background part of the metric $g_{0,AB}$ and $\gamma_{1,\bar{A}\bar{B}}$ denotes pullback from the fluctuated part of the metric $\delta g_{AB}$ evaluated to first order. Since moduli stabilization effects act as small perturbations on the tree-level background and fluctuation, they are negligible for our purposes, and we work with the tree-level supergravity fields only (including treating the dilaton as a constant).
The long string action to first order in fluctuations is therefore
\begin{align}
    S_{\textrm F1} \approx \int dt\left(-T_1 \int d\sigma \sqrt{-\gamma_0} - \frac12 T_1 \int d\sigma\, \sqrt{-\gamma_0}\, \gamma_0^{\bar{A}\bar{B}} \delta g_{CD}\partial_{\bar{A}}X^C \partial_{\bar{B}}X^D\right) .
\end{align}

In order to match to worldsheet CFT string amplitudes, we switch to a coordinate where the background metric is flat, which is possible in the approximation that the string is localized to a point in the compact dimensions. 
We define Riemann normal coordinates around the string's position $y_\star$ at the bottom of the warped throat. The coordinate transformation is 
\begin{align}
     y^m = y_\star^m +e^{2A_0(y_\star)}e^{\Omega_0}\Lambda^m_n z^n+\mathcal{O}(z^2)
\end{align}
where $\Lambda_n^m$ is a matrix that diagonalizes $\tilde{g}_{mn}(y_\star)$. This makes the background metric 
\begin{align}
    ds^2_0=e^{2\Omega_0}e^{2A_0(y_\star)}\hat{\eta}_{MN}dx^{M}dx^{N}
\end{align}
to order $z^2$. Explicitly, we have chosen coordinates $z^m$ such that
\begin{align}
    e^{-2A_0(y_\star)}\tilde{g}_{mn}dy^m dy^n = e^{2\Omega_0}e^{2A_0(y_\star)}(\delta_{mn}+\mathcal{O}(z^2))dz^m dz^n\,.
\end{align}
We write the vector compensator $B_m(y)$ in the Riemann normal coordinates as 
\begin{align}
      B_m dy^m = B_m \frac{dy^m}{dz^n}dz^n=e^{2A_0(y_\star)}e^{\Omega_0}B_m\Lambda^m_n dz^n=e^{2A_0(y_\star)}e^{\Omega_0}\mathcal{B}_n dz^n
\end{align}
The volume modulus couples to fluctuations of the string in both internal and external dimensions:
\begin{align}
    S_{\textrm F1}\supset -\frac{1}{4\pi}\sqrt{\frac{2}{3}}\frac{M_s^2}{M_p}\int d\sigma d\tau\sqrt{-\hat{\gamma}_0}&\left\{\left(\hat{\gamma}_0^{\bar{A}\bar{B}}\hat{\eta}_{\mu\nu}\partial_{\bar{A}}x^\mu \partial_{\bar{B}}x^\nu\right)\left(-1-\frac{1}{2}e^{-2\Omega_0}e^{4A_0(y_\star)}\right)\mathfrak{c}(x)\right. \nonumber\\
    &\left.+2\left(\hat{\gamma}_0^{\bar{A}\bar{B}}\partial_{\bar{A}}x^\mu \partial_{\bar{B}}z^m\right)e^{-\Omega_0}e^{2A_0(y_\star)}\partial_\mu \mathfrak{c}(x)\mathcal{B}_m(0)\nonumber\right.\\
    & \left.+\left(\hat{\gamma}_0^{\bar{A}\bar{B}}\partial_{\bar{A}}z^m \partial_{\bar{B}}z^n\right)\left(\frac{1}{2}e^{-2\Omega_0}e^{4A_0(y_\star)}\delta_{mn}\right)\mathfrak{c}(x)\right\} ,
\end{align}
where $\mathfrak{c}(x)$ is the canonically normalized volume modulus \cite{Frey:2008xw}. $M_s$ is the local string scale in the Einstein frame defined using
\begin{align}
    T_1 =\frac{1}{2\pi}e^{-2\Omega_0}e^{-2A_0(y_\star)}M_s^2 ,
\end{align}
and the 4D Planck mass is
\begin{align}
    M_{p}^2=M_{10}^8 e^{2\Omega_0} \int d^6y\sqrt{{\tilde{g}}}e^{-4A_0}=M_{10}^8 \left( \int d^6y \sqrt{\tilde{g}}\right).
\end{align}
If the string is in a strongly warped region, the dominant coupling is the first term,
\begin{align}
    S_{\textrm F1}\supset \frac{1}{2\pi\sqrt{6}}\frac{M_s^2}{M_{p}}\int dt\int d\sigma\sqrt{-\hat{\gamma}_0}\left(\hat{\gamma}_0^{\bar{A}\bar{B}}\hat{\eta}_{\mu\nu}\partial_{\bar{A}}x^\mu \partial_{\bar{B}}x^\nu\right)\mathfrak{c}(x). 
\end{align}
This is precisely $1/\sqrt{6}$ times the trace of the coupling to gravitons. As an aside, there is an additional diffeomorphism at first order in the metric fluctuations that appears to change these couplings; however, changes due to these diffeomorphisms can be removed using integration by parts on the above action.

\bibliographystyle{JHEP} 
\bibliography{ref.bib}

\providecommand{\href}[2]{#2}\begingroup\raggedright\begin{thebibliography}{10}

\bibitem{Planck:2018vyg}
{\scshape Planck} collaboration, \emph{{Planck 2018 results. VI. Cosmological parameters}}, \href{https://doi.org/10.1051/0004-6361/201833910}{\emph{Astron. Astrophys.} {\bfseries 641} (2020) A6} [\href{https://arxiv.org/abs/1807.06209}{{\ttfamily 1807.06209}}].

\bibitem{Padmanabhan:2005es}
N.~Padmanabhan and D.P.~Finkbeiner, \emph{{Detecting dark matter annihilation with CMB polarization: Signatures and experimental prospects}}, \href{https://doi.org/10.1103/PhysRevD.72.023508}{\emph{Phys. Rev. D} {\bfseries 72} (2005) 023508} [\href{https://arxiv.org/abs/astro-ph/0503486}{{\ttfamily astro-ph/0503486}}].

\bibitem{Galli:2009zc}
S.~Galli, F.~Iocco, G.~Bertone and A.~Melchiorri, \emph{{CMB constraints on Dark Matter models with large annihilation cross-section}}, \href{https://doi.org/10.1103/PhysRevD.80.023505}{\emph{Phys. Rev. D} {\bfseries 80} (2009) 023505} [\href{https://arxiv.org/abs/0905.0003}{{\ttfamily 0905.0003}}].

\bibitem{Slatyer:2009yq}
T.R.~Slatyer, N.~Padmanabhan and D.P.~Finkbeiner, \emph{{CMB Constraints on WIMP Annihilation: Energy Absorption During the Recombination Epoch}}, \href{https://doi.org/10.1103/PhysRevD.80.043526}{\emph{Phys. Rev. D} {\bfseries 80} (2009) 043526} [\href{https://arxiv.org/abs/0906.1197}{{\ttfamily 0906.1197}}].

\bibitem{Slatyer:2012yq}
T.R.~Slatyer, \emph{{Energy Injection And Absorption In The Cosmic Dark Ages}}, \href{https://doi.org/10.1103/PhysRevD.87.123513}{\emph{Phys. Rev. D} {\bfseries 87} (2013) 123513} [\href{https://arxiv.org/abs/1211.0283}{{\ttfamily 1211.0283}}].

\bibitem{Poulin:2016nat}
V.~Poulin, P.D.~Serpico and J.~Lesgourgues, \emph{{A fresh look at linear cosmological constraints on a decaying dark matter component}}, \href{https://doi.org/10.1088/1475-7516/2016/08/036}{\emph{JCAP} {\bfseries 08} (2016) 036} [\href{https://arxiv.org/abs/1606.02073}{{\ttfamily 1606.02073}}].

\bibitem{Simon:2022ftd}
T.~Simon, G.~Franco~Abell{\'a}n, P.~Du, V.~Poulin and Y.~Tsai, \emph{{Constraining decaying dark matter with BOSS data and the effective field theory of large-scale structures}}, \href{https://doi.org/10.1103/PhysRevD.106.023516}{\emph{Phys. Rev. D} {\bfseries 106} (2022) 023516} [\href{https://arxiv.org/abs/2203.07440}{{\ttfamily 2203.07440}}].

\bibitem{Liu:2023fgu}
H.~Liu, W.~Qin, G.W.~Ridgway and T.R.~Slatyer, \emph{{Exotic energy injection in the early Universe. I. A novel treatment for low-energy electrons and photons}}, \href{https://doi.org/10.1103/PhysRevD.108.043530}{\emph{Phys. Rev. D} {\bfseries 108} (2023) 043530} [\href{https://arxiv.org/abs/2303.07366}{{\ttfamily 2303.07366}}].

\bibitem{Xu:2024vdn}
C.~Xu, W.~Qin and T.R.~Slatyer, \emph{{CMB limits on decaying dark matter beyond the ionization threshold}}, \href{https://doi.org/10.1103/PhysRevD.110.123529}{\emph{Phys. Rev. D} {\bfseries 110} (2024) 123529} [\href{https://arxiv.org/abs/2408.13305}{{\ttfamily 2408.13305}}].

\bibitem{Cicoli:2023opf}
M.~Cicoli, J.P.~Conlon, A.~Maharana, S.~Parameswaran, F.~Quevedo and I.~Zavala, \emph{{String cosmology: From the early universe to today}}, \href{https://doi.org/10.1016/j.physrep.2024.01.002}{\emph{Phys. Rept.} {\bfseries 1059} (2024) 1} [\href{https://arxiv.org/abs/2303.04819}{{\ttfamily 2303.04819}}].

\bibitem{Coughlan:1983ci}
G.D.~Coughlan, W.~Fischler, E.W.~Kolb, S.~Raby and G.G.~Ross, \emph{{Cosmological Problems for the Polonyi Potential}}, \href{https://doi.org/10.1016/0370-2693(83)91091-2}{\emph{Phys. Lett. B} {\bfseries 131} (1983) 59}.

\bibitem{Banks:1993en}
T.~Banks, D.B.~Kaplan and A.E.~Nelson, \emph{{Cosmological implications of dynamical supersymmetry breaking}}, \href{https://doi.org/10.1103/PhysRevD.49.779}{\emph{Phys. Rev. D} {\bfseries 49} (1994) 779} [\href{https://arxiv.org/abs/hep-ph/9308292}{{\ttfamily hep-ph/9308292}}].

\bibitem{deCarlos:1993wie}
B.~de~Carlos, J.A.~Casas, F.~Quevedo and E.~Roulet, \emph{{Model independent properties and cosmological implications of the dilaton and moduli sectors of 4-d strings}}, \href{https://doi.org/10.1016/0370-2693(93)91538-X}{\emph{Phys. Lett. B} {\bfseries 318} (1993) 447} [\href{https://arxiv.org/abs/hep-ph/9308325}{{\ttfamily hep-ph/9308325}}].

\bibitem{Kawasaki:2000en}
M.~Kawasaki, K.~Kohri and N.~Sugiyama, \emph{{MeV scale reheating temperature and thermalization of neutrino background}}, \href{https://doi.org/10.1103/PhysRevD.62.023506}{\emph{Phys. Rev. D} {\bfseries 62} (2000) 023506} [\href{https://arxiv.org/abs/astro-ph/0002127}{{\ttfamily astro-ph/0002127}}].

\bibitem{Preskill:1982cy}
J.~Preskill, M.B.~Wise and F.~Wilczek, \emph{{Cosmology of the Invisible Axion}}, \href{https://doi.org/10.1016/0370-2693(83)90637-8}{\emph{Phys. Lett. B} {\bfseries 120} (1983) 127}.

\bibitem{Balazs:2022tjl}
C.~Bal{\'a}zs et~al., \emph{{Cosmological constraints on decaying axion-like particles: a global analysis}}, \href{https://doi.org/10.1088/1475-7516/2022/12/027}{\emph{JCAP} {\bfseries 12} (2022) 027} [\href{https://arxiv.org/abs/2205.13549}{{\ttfamily 2205.13549}}].

\bibitem{Langhoff:2022bij}
K.~Langhoff, N.J.~Outmezguine and N.L.~Rodd, \emph{{Irreducible Axion Background}}, \href{https://doi.org/10.1103/PhysRevLett.129.241101}{\emph{Phys. Rev. Lett.} {\bfseries 129} (2022) 241101} [\href{https://arxiv.org/abs/2209.06216}{{\ttfamily 2209.06216}}].

\bibitem{Ghiglieri:2015nfa}
J.~Ghiglieri and M.~Laine, \emph{{Gravitational wave background from Standard Model physics: Qualitative features}}, \href{https://doi.org/10.1088/1475-7516/2015/07/022}{\emph{JCAP} {\bfseries 07} (2015) 022} [\href{https://arxiv.org/abs/1504.02569}{{\ttfamily 1504.02569}}].

\bibitem{Ghiglieri:2020mhm}
J.~Ghiglieri, G.~Jackson, M.~Laine and Y.~Zhu, \emph{{Gravitational wave background from Standard Model physics: Complete leading order}}, \href{https://doi.org/10.1007/JHEP07(2020)092}{\emph{JHEP} {\bfseries 07} (2020) 092} [\href{https://arxiv.org/abs/2004.11392}{{\ttfamily 2004.11392}}].

\bibitem{Ringwald:2020ist}
A.~Ringwald, J.~Sch{\"u}tte-Engel and C.~Tamarit, \emph{{Gravitational Waves as a Big Bang Thermometer}}, \href{https://doi.org/10.1088/1475-7516/2021/03/054}{\emph{JCAP} {\bfseries 03} (2021) 054} [\href{https://arxiv.org/abs/2011.04731}{{\ttfamily 2011.04731}}].

\bibitem{Cheng:2025cmb}
H.~Cheng, Z.~Yin, E.~Di~Valentino, D.J.E.~Marsh and L.~Visinelli, \emph{{Constraining exotic high-$z$ reionization histories with Gaussian processes and the Cosmic Microwave Background}},  \href{https://arxiv.org/abs/2506.19096}{{\ttfamily 2506.19096}}.

\bibitem{Yin:2025amn}
Z.~Yin, H.~Cheng, E.~Di~Valentino, N.~Gendler, D.J.E.~Marsh and L.~Visinelli, \emph{{Constraining the axiverse with reionization}},  \href{https://arxiv.org/abs/2507.03535}{{\ttfamily 2507.03535}}.

\bibitem{Frey:2005jk}
A.R.~Frey, A.~Mazumdar and R.C.~Myers, \emph{{Stringy effects during inflation and reheating}}, \href{https://doi.org/10.1103/PhysRevD.73.026003}{\emph{Phys. Rev. D} {\bfseries 73} (2006) 026003} [\href{https://arxiv.org/abs/hep-th/0508139}{{\ttfamily hep-th/0508139}}].

\bibitem{Frey:2023khe}
A.R.~Frey, R.~Mahanta, A.~Maharana, F.~Muia, F.~Quevedo and G.~Villa, \emph{{String thermodynamics in and out of equilibrium: Boltzmann equations and random walks}}, \href{https://doi.org/10.1007/JHEP03(2024)112}{\emph{JHEP} {\bfseries 03} (2024) 112} [\href{https://arxiv.org/abs/2310.11494}{{\ttfamily 2310.11494}}].

\bibitem{Frey:2024jqy}
A.R.~Frey, R.~Mahanta, A.~Maharana, F.~Quevedo and G.~Villa, \emph{{Gravitational waves from high temperature strings}}, \href{https://doi.org/10.1007/JHEP12(2024)174}{\emph{JHEP} {\bfseries 12} (2024) 174} [\href{https://arxiv.org/abs/2408.13803}{{\ttfamily 2408.13803}}].

\bibitem{Kapusta:1991qp}
J.I.~Kapusta, P.~Lichard and D.~Seibert, \emph{{High-energy photons from quark - gluon plasma versus hot hadronic gas}}, \href{https://doi.org/10.1103/PhysRevD.47.4171}{\emph{Phys. Rev. D} {\bfseries 44} (1991) 2774}.

\bibitem{Arnold:2001ms}
P.B.~Arnold, G.D.~Moore and L.G.~Yaffe, \emph{{Photon emission from quark gluon plasma: Complete leading order results}}, \href{https://doi.org/10.1088/1126-6708/2001/12/009}{\emph{JHEP} {\bfseries 12} (2001) 009} [\href{https://arxiv.org/abs/hep-ph/0111107}{{\ttfamily hep-ph/0111107}}].

\bibitem{Bolz:2000fu}
M.~Bolz, A.~Brandenburg and W.~Buchmuller, \emph{{Thermal production of gravitinos}}, \href{https://doi.org/10.1016/S0550-3213(01)00132-8}{\emph{Nucl. Phys. B} {\bfseries 606} (2001) 518} [\href{https://arxiv.org/abs/hep-ph/0012052}{{\ttfamily hep-ph/0012052}}].

\bibitem{Braaten:1991dd}
E.~Braaten and T.C.~Yuan, \emph{{Calculation of screening in a hot plasma}}, \href{https://doi.org/10.1103/PhysRevLett.66.2183}{\emph{Phys. Rev. Lett.} {\bfseries 66} (1991) 2183}.

\bibitem{Graf:2010tv}
P.~Graf and F.D.~Steffen, \emph{{Thermal axion production in the primordial quark-gluon plasma}}, \href{https://doi.org/10.1103/PhysRevD.83.075011}{\emph{Phys. Rev. D} {\bfseries 83} (2011) 075011} [\href{https://arxiv.org/abs/1008.4528}{{\ttfamily 1008.4528}}].

\bibitem{Graf:2012hb}
P.~Graf and F.D.~Steffen, \emph{{Axions and saxions from the primordial supersymmetric plasma and extra radiation signatures}}, \href{https://doi.org/10.1088/1475-7516/2013/02/018}{\emph{JCAP} {\bfseries 02} (2013) 018} [\href{https://arxiv.org/abs/1208.2951}{{\ttfamily 1208.2951}}].

\bibitem{Besak:2012qm}
D.~Besak and D.~Bodeker, \emph{{Thermal production of ultrarelativistic right-handed neutrinos: Complete leading-order results}}, \href{https://doi.org/10.1088/1475-7516/2012/03/029}{\emph{JCAP} {\bfseries 03} (2012) 029} [\href{https://arxiv.org/abs/1202.1288}{{\ttfamily 1202.1288}}].

\bibitem{Braaten:1989mz}
E.~Braaten and R.D.~Pisarski, \emph{{Soft Amplitudes in Hot Gauge Theories: A General Analysis}}, \href{https://doi.org/10.1016/0550-3213(90)90508-B}{\emph{Nucl. Phys. B} {\bfseries 337} (1990) 569}.

\bibitem{Chakraborty:2026hob}
D.~Chakraborty and A.R.~Kamal, \emph{{Fire at the Tip of the Throat: Hagedorn Phase after brane-antibrane inflation?}},  \href{https://arxiv.org/abs/2606.08767}{{\ttfamily 2606.08767}}.

\bibitem{Deo:1988jj}
N.~Deo, S.~Jain and C.-I.~Tan, \emph{{Strings at High-energy Densities and Complex Temperature}}, \href{https://doi.org/10.1016/0370-2693(89)90024-5}{\emph{Phys. Lett. B} {\bfseries 220} (1989) 125}.

\bibitem{Deo:1989bv}
N.~Deo, S.~Jain and C.-I.~Tan, \emph{{String distributions above the Hagedorn energy density}}, \href{https://doi.org/10.1103/PhysRevD.40.2626}{\emph{Phys. Rev. D} {\bfseries 40} (1989) 2626}.

\bibitem{Kachru:2003aw}
S.~Kachru, R.~Kallosh, A.D.~Linde and S.P.~Trivedi, \emph{{De Sitter vacua in string theory}}, \href{https://doi.org/10.1103/PhysRevD.68.046005}{\emph{Phys. Rev. D} {\bfseries 68} (2003) 046005} [\href{https://arxiv.org/abs/hep-th/0301240}{{\ttfamily hep-th/0301240}}].

\bibitem{Balasubramanian:2005zx}
V.~Balasubramanian, P.~Berglund, J.P.~Conlon and F.~Quevedo, \emph{{Systematics of moduli stabilisation in Calabi-Yau flux compactifications}}, \href{https://doi.org/10.1088/1126-6708/2005/03/007}{\emph{JHEP} {\bfseries 03} (2005) 007} [\href{https://arxiv.org/abs/hep-th/0502058}{{\ttfamily hep-th/0502058}}].

\bibitem{Agarwal:2025rqd}
N.~Agarwal, A.R.~Frey and B.~Underwood, \emph{{Toward an effective theory of the volume modulus}}, \href{https://doi.org/10.1007/JHEP01(2026)136}{\emph{JHEP} {\bfseries 01} (2026) 136} [\href{https://arxiv.org/abs/2509.18419}{{\ttfamily 2509.18419}}].

\bibitem{Kawasaki:1997ah}
M.~Kawasaki and T.~Yanagida, \emph{{Constraint on cosmic density of the string moduli field in gauge mediated supersymmetry breaking theories}}, \href{https://doi.org/10.1016/S0370-2693(97)00282-7}{\emph{Phys. Lett. B} {\bfseries 399} (1997) 45} [\href{https://arxiv.org/abs/hep-ph/9701346}{{\ttfamily hep-ph/9701346}}].

\bibitem{Adelberger:2003zx}
E.G.~Adelberger, B.R.~Heckel and A.E.~Nelson, \emph{{Tests of the gravitational inverse square law}}, \href{https://doi.org/10.1146/annurev.nucl.53.041002.110503}{\emph{Ann. Rev. Nucl. Part. Sci.} {\bfseries 53} (2003) 77} [\href{https://arxiv.org/abs/hep-ph/0307284}{{\ttfamily hep-ph/0307284}}].

\bibitem{Kapner:2006si}
D.J.~Kapner, T.S.~Cook, E.G.~Adelberger, J.H.~Gundlach, B.R.~Heckel, C.D.~Hoyle et~al., \emph{{Tests of the gravitational inverse-square law below the dark-energy length scale}}, \href{https://doi.org/10.1103/PhysRevLett.98.021101}{\emph{Phys. Rev. Lett.} {\bfseries 98} (2007) 021101} [\href{https://arxiv.org/abs/hep-ph/0611184}{{\ttfamily hep-ph/0611184}}].

\bibitem{Arza:2026rsl}
A.~Arza et~al., \emph{{The COSMIC WISPers White Paper: The physics case for Weakly Interacting Slim Particles}},  \href{https://arxiv.org/abs/2603.03433}{{\ttfamily 2603.03433}}.

\bibitem{SimonsObservatory:2025wwn}
{\scshape Simons Observatory} collaboration, \emph{{The Simons Observatory: science goals and forecasts for the enhanced Large Aperture Telescope}}, \href{https://doi.org/10.1088/1475-7516/2025/08/034}{\emph{JCAP} {\bfseries 08} (2025) 034} [\href{https://arxiv.org/abs/2503.00636}{{\ttfamily 2503.00636}}].

\bibitem{LiteBIRD:2024dbi}
{\scshape LiteBIRD} collaboration, \emph{{LiteBIRD science goals and forecasts. Mapping the hot gas in the Universe}}, \href{https://doi.org/10.1088/1475-7516/2024/12/026}{\emph{JCAP} {\bfseries 12} (2024) 026} [\href{https://arxiv.org/abs/2407.17555}{{\ttfamily 2407.17555}}].

\bibitem{NASAPICO:2019thw}
{\scshape NASA PICO} collaboration, \emph{{PICO: Probe of Inflation and Cosmic Origins}},  \href{https://arxiv.org/abs/1902.10541}{{\ttfamily 1902.10541}}.

\bibitem{CMB-S4:2016ple}
{\scshape CMB-S4} collaboration, \emph{{CMB-S4 Science Book, First Edition}} (10, 2016), \href{https://doi.org/10.2172/1352047}{10.2172/1352047}, [\href{https://arxiv.org/abs/1610.02743}{{\ttfamily 1610.02743}}].

\bibitem{DAmico:2018sxd}
G.~D'Amico, P.~Panci and A.~Strumia, \emph{{Bounds on Dark Matter annihilations from 21 cm data}}, \href{https://doi.org/10.1103/PhysRevLett.121.011103}{\emph{Phys. Rev. Lett.} {\bfseries 121} (2018) 011103} [\href{https://arxiv.org/abs/1803.03629}{{\ttfamily 1803.03629}}].

\bibitem{Mitridate:2018iag}
A.~Mitridate and A.~Podo, \emph{{Bounds on Dark Matter decay from 21 cm line}}, \href{https://doi.org/10.1088/1475-7516/2018/05/069}{\emph{JCAP} {\bfseries 05} (2018) 069} [\href{https://arxiv.org/abs/1803.11169}{{\ttfamily 1803.11169}}].

\bibitem{Sun:2023acy}
Y.~Sun, J.W.~Foster, H.~Liu, J.B.~Mu{\~n}oz and T.R.~Slatyer, \emph{{Inhomogeneous energy injection in the 21-cm power spectrum: Sensitivity to dark matter decay}}, \href{https://doi.org/10.1103/PhysRevD.111.043015}{\emph{Phys. Rev. D} {\bfseries 111} (2025) 043015} [\href{https://arxiv.org/abs/2312.11608}{{\ttfamily 2312.11608}}].

\bibitem{Zhao:2025ddy}
M.-L.~Zhao, Y.~Shao, S.~Wang and X.~Zhang, \emph{{Prospects for probing dark matter particles and primordial black holes with the Square Kilometre Array using the 21 cm power spectrum at cosmic dawn*}}, \href{https://doi.org/10.1088/1674-1137/ae1375}{\emph{Chin. Phys. C} {\bfseries 50} (2026) 025101} [\href{https://arxiv.org/abs/2507.02651}{{\ttfamily 2507.02651}}].

\bibitem{Zeldovich:1969ff}
Y.B.~Zeldovich and R.A.~Sunyaev, \emph{{The Interaction of Matter and Radiation in a Hot-Model Universe}}, \href{https://doi.org/10.1007/BF00661821}{\emph{Astrophys. Space Sci.} {\bfseries 4} (1969) 301}.

\bibitem{Chluba:2011hw}
J.~Chluba and R.A.~Sunyaev, \emph{{The evolution of CMB spectral distortions in the early Universe}}, \href{https://doi.org/10.1111/j.1365-2966.2011.19786.x}{\emph{Mon. Not. Roy. Astron. Soc.} {\bfseries 419} (2012) 1294} [\href{https://arxiv.org/abs/1109.6552}{{\ttfamily 1109.6552}}].

\bibitem{Essig:2013goa}
R.~Essig, E.~Kuflik, S.D.~McDermott, T.~Volansky and K.M.~Zurek, \emph{{Constraining Light Dark Matter with Diffuse X-Ray and Gamma-Ray Observations}}, \href{https://doi.org/10.1007/JHEP11(2013)193}{\emph{JHEP} {\bfseries 11} (2013) 193} [\href{https://arxiv.org/abs/1309.4091}{{\ttfamily 1309.4091}}].

\bibitem{Boyarsky:2006zi}
A.~Boyarsky, A.~Neronov, O.~Ruchayskiy and M.~Shaposhnikov, \emph{{Restrictions on parameters of sterile neutrino dark matter from observations of galaxy clusters}}, \href{https://doi.org/10.1103/PhysRevD.74.103506}{\emph{Phys. Rev. D} {\bfseries 74} (2006) 103506} [\href{https://arxiv.org/abs/astro-ph/0603368}{{\ttfamily astro-ph/0603368}}].

\bibitem{Agarwal:2026mtm}
N.~Agarwal, A.R.~Frey, R.~Mahanta and E.~McDonough, \emph{{Quadratic Axion Couplings in String Theory}},  \href{https://arxiv.org/abs/2607.27190}{{\ttfamily 2607.27190}}.

\bibitem{Dasgupta:1999ss}
K.~Dasgupta, G.~Rajesh and S.~Sethi, \emph{{M theory, orientifolds and G - flux}}, \href{https://doi.org/10.1088/1126-6708/1999/08/023}{\emph{JHEP} {\bfseries 08} (1999) 023} [\href{https://arxiv.org/abs/hep-th/9908088}{{\ttfamily hep-th/9908088}}].

\bibitem{Greene:2000gh}
B.R.~Greene, K.~Schalm and G.~Shiu, \emph{{Warped compactifications in M and F theory}}, \href{https://doi.org/10.1016/S0550-3213(00)00400-4}{\emph{Nucl. Phys. B} {\bfseries 584} (2000) 480} [\href{https://arxiv.org/abs/hep-th/0004103}{{\ttfamily hep-th/0004103}}].

\bibitem{Giddings:2001yu}
S.B.~Giddings, S.~Kachru and J.~Polchinski, \emph{{Hierarchies from fluxes in string compactifications}}, \href{https://doi.org/10.1103/PhysRevD.66.106006}{\emph{Phys. Rev. D} {\bfseries 66} (2002) 106006} [\href{https://arxiv.org/abs/hep-th/0105097}{{\ttfamily hep-th/0105097}}].

\bibitem{Becker:1996gj}
K.~Becker and M.~Becker, \emph{{M theory on eight manifolds}}, \href{https://doi.org/10.1016/0550-3213(96)00367-7}{\emph{Nucl. Phys. B} {\bfseries 477} (1996) 155} [\href{https://arxiv.org/abs/hep-th/9605053}{{\ttfamily hep-th/9605053}}].

\bibitem{Jackson:2004zg}
M.G.~Jackson, N.T.~Jones and J.~Polchinski, \emph{{Collisions of cosmic F and D-strings}}, \href{https://doi.org/10.1088/1126-6708/2005/10/013}{\emph{JHEP} {\bfseries 10} (2005) 013} [\href{https://arxiv.org/abs/hep-th/0405229}{{\ttfamily hep-th/0405229}}].

\bibitem{PandoZayas:2003jr}
L.A.~Pando~Zayas and D.~Vaman, \emph{{Hadronic density of states from string theory}}, \href{https://doi.org/10.1103/PhysRevLett.91.111602}{\emph{Phys. Rev. Lett.} {\bfseries 91} (2003) 111602} [\href{https://arxiv.org/abs/hep-th/0306107}{{\ttfamily hep-th/0306107}}].

\bibitem{Canneti:2024iyn}
T.~Canneti, \emph{{On the asymptotic density of states in solvable models of strings}}, \href{https://doi.org/10.1007/JHEP12(2024)043}{\emph{JHEP} {\bfseries 12} (2024) 043} [\href{https://arxiv.org/abs/2406.08405}{{\ttfamily 2406.08405}}].

\bibitem{Lowe:1994nm}
D.A.~Lowe and L.~Thorlacius, \emph{{Hot string soup}}, \href{https://doi.org/10.1103/PhysRevD.51.665}{\emph{Phys. Rev. D} {\bfseries 51} (1995) 665} [\href{https://arxiv.org/abs/hep-th/9408134}{{\ttfamily hep-th/9408134}}].

\bibitem{Lee:1997iz}
S.~Lee and L.~Thorlacius, \emph{{Strings and D-branes at high temperature}}, \href{https://doi.org/10.1016/S0370-2693(97)01105-2}{\emph{Phys. Lett. B} {\bfseries 413} (1997) 303} [\href{https://arxiv.org/abs/hep-th/9707167}{{\ttfamily hep-th/9707167}}].

\bibitem{Copeland:1998na}
E.J.~Copeland, T.W.B.~Kibble and D.A.~Steer, \emph{{The Evolution of a network of cosmic string loops}}, \href{https://doi.org/10.1103/PhysRevD.58.043508}{\emph{Phys. Rev. D} {\bfseries 58} (1998) 043508} [\href{https://arxiv.org/abs/hep-ph/9803414}{{\ttfamily hep-ph/9803414}}].

\bibitem{Giddings:2005ff}
S.B.~Giddings and A.~Maharana, \emph{{Dynamics of warped compactifications and the shape of the warped landscape}}, \href{https://doi.org/10.1103/PhysRevD.73.126003}{\emph{Phys. Rev. D} {\bfseries 73} (2006) 126003} [\href{https://arxiv.org/abs/hep-th/0507158}{{\ttfamily hep-th/0507158}}].

\bibitem{Frey:2008xw}
A.R.~Frey, G.~Torroba, B.~Underwood and M.R.~Douglas, \emph{{The Universal Kahler Modulus in Warped Compactifications}}, \href{https://doi.org/10.1088/1126-6708/2009/01/036}{\emph{JHEP} {\bfseries 01} (2009) 036} [\href{https://arxiv.org/abs/0810.5768}{{\ttfamily 0810.5768}}].

\bibitem{weinberg2015lectures}
S.~Weinberg, \emph{Lectures on quantum mechanics}, Cambridge University Press (2015).

\end{thebibliography}\endgroup
\end{document}